\documentclass[prd,nofootinbib,]{revtex4}
\usepackage{latexsym}
\usepackage{amsthm}
\usepackage{amsmath}
\usepackage{graphicx}  
\usepackage{bm}        
\usepackage{amssymb}   
\usepackage{hyperref}
\usepackage{dsfont}
\newcommand{\gsim}{\lower.7ex\hbox{$\;\stackrel{\textstyle>}{\sim}\;$}}
\newcommand{\lsim}{\lower.7ex\hbox{$\;\stackrel{\textstyle<}{\sim}\;$}}

\newcommand{\bef}{\begin{figure}[htbp]\begin{center}}
\newcommand{\eef}{\end{center}\end{figure}}

\newcommand{\be}{\begin{equation}}
\newcommand{\ee}{\end{equation}}
\newcommand{\beq}{\begin{eqnarray}}
\newcommand{\eeq}{\end{eqnarray}}
\newcommand{\bea}{\begin{eqnarray}}
\newcommand{\eea}{\end{eqnarray}}
\newcommand{\beqn}{\begin{eqnarray}}
\newcommand{\eeqn}{\end{eqnarray}}
\newcommand{\rd}{\mathrm{d}}

\def\pa{\partial}

\def\dd{\!\cdot \!}

\def\nn{\nonumber}

\begin{document}

\title{Snyder Momentum Space in Relative Locality}

\author{Andrzej Banburski, Laurent Freidel}
\affiliation{Perimeter Institute for Theoretical Physics, Waterloo, ON N2L 2Y5, Canada}
\date{\today}

\begin{abstract}
The standard approaches of phenomenology of Quantum Gravity have  usually explicitly violated  Lorentz invariance, either in the dispersion relation or in the addition rule for momenta. We investigate whether it is possible in 3+1 dimensions to have a non local deformation that preserves fully Lorentz invariance,
as it is the case in  2+1D Quantum Gravity. We answer positively to this question and show for the first time how to construct a  homogeneously curved momentum space preserving the full action of the Lorentz group in  dimension 4  and higher, despite relaxing locality. We study the property of this relative locality deformation and show that  this space leads to a noncommutativity
related to  Snyder spacetime.
\end{abstract}
\maketitle


\section{Introduction}
The study of Quantum Gravity phenomenology in the form of \emph{Doubly Special Relativity} \cite{AmelinoCamelia:2000mn, AmelinoCamelia:2000ge, Magueijo:2002am} or noncommutative field theories \cite{Majid:1999tc, Doplicher:1994tu} has been mostly based  on understanding the effects of violating Lorentz invariance. This has been motivated by the expectation that the inclusion of a minimal length scale in Quantum Gravity should ultimately make Lorentz symmetry emergent in low-energy regime. 
Even the more covariant case  of $\kappa$-Poincar\'e, which carries a deformed  action of the Lorentz group on one particle states, does not respect the symmetry on multiparticle states (see \cite{Kowalski1} for a review) .
However, the known example of 2+1 -- dimensional Quantum Gravity 
\cite{Witten:1988hc} coupled to point particles does preserve full Lorentz invariance. It is thus reasonable to suspect that we should study Lorentz invariant Quantum Gravity induced deformations to low-energy physics.

Another lesson of 2+1D Quantum Gravity coupled to point particles is that it can be realized as a theory with a non-trivial momentum space geometry \cite{FL, Freidel:2003sp, Schroers:2007ey}. 
More generally, exact Lorentz invariance is one of the key symmetries in local field theory.
It is of fundamental interest to know whether it can be exactly implemented even in the context where locality is relaxed. 
Moreover, a very recent result shows that in the context of string theory, which preserves Lorentz symmetry, momentum space is generically curved \cite{Born}. Given all these clues, we investigate whether it is possible to have a non-trivial 3+1D deformation of relativistic dynamics of point particles preserving Lorentz invariance. We work in the framework of \emph{Relative Locality} (RL) \cite{AmelinoCamelia:2011bm}, where momentum space is taken as a primary object, and spacetime is emergent. The relaxation of locality in this approach is a response to the impossibility of recovering exact locality in Doubly Special Relativity \cite{Hossenfelder:2010tm} for interactions distant from the observer. The basic motivation behind Relative Locality is that our interaction with the physical world is fundamentally that of measuring energy and direction of incoming photons -- we never directly probe spacetime itself. Hence, there is no \emph{a priori} reason to assume that momentum space has to be flat and that locality must be absolute.

Within this framework, we find that indeed it is possible to have a non-trivial momentum space with undeformed action of Lorentz group. The construction is based on a homogenously curved momentum space -- de Sitter, with an addition rule preserving the metric structure. We require that the addition rule is the same before and after a Lorentz boost -- this condition excludes the popular associative and non-commutative $\kappa$-Poincar\'e  addition rule \cite{Gubitosi:2011ej}. The addition rule we obtain preserves full Lorentz symmetry, but is necessarily nonassociative, which is directly related to the curvature of the momentum space. This seems to be a general consequence of demanding exact Lorentz invariance.
We study the properties of the emergent spacetime structure and find that it is related to the first ever studied quantized spacetime, proposed by Snyder in 1946 \cite{Snyder:1946qz}. The QFT on the momentum space of Snyder spacetime was studied previously for example in \cite{Golfand0, Golfand1, Golfand2}. We then go on to study closed loop processes on the momentum space and come to the surprising conclusion that it seems impossible to synchronize simultaneously more than two clocks in an observer-independent fashion in this theory. 
During the final completion of this work we became aware of \cite{ACetal}, which investigates similar structures.


\section{Dynamics of Relative Locality}
We will start by reviewing the basic principles of relative locality \cite{AmelinoCamelia:2011bm, Kowalski1}. 
According to these principles, if momentum space $\mathcal{P}$ is assumed to
have a nontrivial geometry,  there is no meaning to a universal spacetime. 
Instead, the phase space associated with each particle is isomorphic to the cotangent bundle $T^*_p\mathcal{P}$
and   the spacetimes are cotangent planes to points
on the momentum space.

We assume that momentum space is a metric manifold with a preferred point, the origin that corresponds to zero energy and momenta. We define  the rest energy (rest mass for a massive particle) to be given by the geodesic distance from origin to point $p$ as 
\begin{equation}
D^2(p) = m^2,\label{eq:geodesicmass}
\end{equation}
which also gives us the dispersion relation.
In order to describe interactions of particles, we need the notion of addition of momenta. Since we want to work with nontrivial geometry of momentum space,
we assume that this is in general a non-linear composition law. The addition rule is defined to be a $C^{\infty}$ map
\begin{equation}
\begin{split}
\oplus &: \mathcal{P}\times\mathcal{P} \rightarrow \mathcal{P} \\
&(p,q)\mapsto p\oplus q
\end{split}
\end{equation}
such that it has an identity $0$ 
\begin{equation}
0\oplus p = p\oplus 0 = p,
\end{equation}
and an inverse $\ominus$
\begin{equation}
\begin{split}
&\ominus : \mathcal{P}\rightarrow\mathcal{P}, \\
&\ominus p \oplus p = p\ominus p = 0 .
\end{split}
\end{equation}
We also introduce the notion of  left translation operator $L_p$ and right translation operator $R_p$, such that
\begin{equation}
L_p(q) \equiv p\oplus q, \ \ \ R_p(q)\equiv q\oplus p.
\end{equation}
We assume that both $L_{p}$ and $R_{p}$ are invertible maps.
If in addition we have that $L_p^{-1} = L_{\ominus p}$ we say that the addition is left invertible, similarly it is right invertible if $R_p^{-1} = R_{\ominus p}$.

It will be useful for later to define transport operators as derivatives of the translation operators $L_p$ and $R_q$.
The left handed transport operator is defined as
\begin{equation}
\left(U^q_{p\oplus q}\right)^\nu_\mu =   \frac{\partial\left(p\oplus q\right)_\mu}{\partial q_\nu}.\label{eq:lefttr}
\end{equation}
Similarly we define the right-handed transport as
\begin{equation}
\left(V^p_{p\oplus q}\right)^\nu_\mu =  \frac{\partial\left(p\oplus q\right)_\mu}{\partial p_\nu}.\label{eq:righttr}
\end{equation}
We can also define a derivative of the inverse, which we will denote by
\begin{equation}
\left(I^p\right)^\nu_\mu = \textnormal{d}_p \ominus = \frac{\partial\left(\ominus p\right)_\mu}{\partial p_\nu}.
\end{equation}
A useful result  is that the three operators $U$, $V$ and $I$ are not independent (if the addition is left invertible), but are related by
\begin{equation}
-V^{\ominus p}_0 I^p = U^p_0. \label{eq:UVrelation}
\end{equation}
The proof of this is as follows. Consider the expression 
\begin{equation}
0 = \textnormal{d}_p\left(\ominus p\oplus\left(p\oplus q\right)\right)  = U^{p\oplus q}_q V^p_{p\oplus q} +
V^{\ominus p}_q I^p \label{eq:dpq0}
\end{equation}
where we used Leibnitz rule. Setting $q=0$ we get (\ref{eq:UVrelation}). 

It is important to note that the addition rule is coordinate dependant, and so if we know it in some set of coordinates, say $P$,
and we have $F_\mu(P) = p_\mu$, then the addition in new coordinates $p$ is related to the old one by
\begin{equation}
p\hat{\oplus}q = F\left(F^{-1}(p)\oplus F^{-1}(q)\right).
\end{equation}
We do not assume that the rule $\oplus$ should be linear, commutative or even associative. For more definitions and details on the algebra of momentum addition, see \cite{Freidel:2012gm}. It is useful to discuss here the so-called "soccer-ball problem" \cite{Hossenfelder:2007fy}. In theories with non-linear momentum addition rules, like the DSR, there has been a problem of discussing macroscopic bodies, since naively putting a cut-off of Planck scale on maximum energy and momentum does not allow composite systems to possess energies bigger than that cut-off. Macroscopic bodies however, a soccer ball for example, have mass and energy-momentum many timess bigger than $m_P$. Relative locality has been shown to be free of the problem \cite{AmelinoCamelia:2011uk, Amelino-Camelia:2013zja, Amelino-Camelia:2014gga}, as the non-linear effects in a system of $N$ particles are suppressed by a factor of $\frac{1}{N m_P}$.

The dynamics of interacting point particles in Relative Locality is governed by the following action
\begin{equation}
S = \sum_{\textnormal{worldines}\  J}S^J_{\textnormal{free}} + \sum_{\textnormal{interactions}\  i} S^i_{\textnormal{int}}. \label{eq:action}
\end{equation}
The free part of the action is
\begin{equation}
S_{\textnormal{free}}^J = \int \textnormal{d}s \left(x^\mu_J \dot{p}_\mu^J + \mathcal{N}_J\left(D^2\left(p^{J}\right) - m^2\right) \right),
\end{equation}
where $J$ stands for the $J$-th particle worldline. $ \mathcal{N}_J$ is a Lagrange multiplier for the mass-shell constraint.
The interaction term in the action is
\begin{equation}
S_{\textnormal{int}}^i = z^\mu\mathcal{K}^i(p(0))_\mu,
\end{equation}
where $\mathcal{K}$ is a conservation law for momenta (for example $k^1 \oplus k^2 \ominus k^3 = 0$), evaluated at affine parameter $s=0$. $z^\mu$ are the Lagrange multipliers for the constraint, but since the constraint lives
at the origin of momentum space, we have $z^\mu\in \textnormal{T}^*_0 \mathcal{P}$. $x^\mu$ are variables conjugate to momenta $p_\mu$ and satisfy
\begin{equation}
\left\{p_\mu,x^\nu\right\} = \delta^\mu_\nu,
\end{equation}
which implies that $x^\mu$ are coordinates ot the cotangent space $\textnormal{T}^*_p \mathcal{P}$. Note
that since the momentum space is nonlinear, $x^\mu$ is only defined as a vector in the cotangent plane at $p$, and hence worldlines of particles with different momenta live on different cotangent spaces. The relation between different worldlines is given by the equation of motion obtained by varying the boundary terms in the action:
\begin{equation}
x^\mu_J(0) =\pm z^\nu\frac{\delta\mathcal{K}_\nu}{\delta p^J_\mu},\label{eq:relxz}
\end{equation}
where sign corresponds to an incoming or outgoing particle. We call $z$ the interaction coordinates,
since the interactions take place at the origin of momentum space. The $\frac{\delta\mathcal{K}_\nu}{\delta p^J_\mu}$ is a transport operator that transports covectors from $\textnormal{T}^*_0 \mathcal{P}$ to $\textnormal{T}^*_{p^J} \mathcal{P}$, so the interactions provide us with a natural path for transporting from the spacetime at momentum $p^J$ to the interaction plane. 


\section{Lorentz symmetry in Relative Locality and de Sitter}

Lorentz symmetry is a key ingredient of all of our fundamental physical theories. 
One of the pressing questions in the context of Relative Locality is whether it is possible to preserve exact Lorentz invariance 
while allowing momentum space to be curved?
We now want to construct the most general geometry of momentum space invariant under Lorentz transformations.
Let us start by considering that the Lorentz action could be deformed.
Now an important point worth emphasizing is that, since the Lorentz group is simple,  it is impossible to deform.
 More precisely, any deformation of its Lie algebra that is still a Lie algebra can be recast in terms of a redefinition of its generators.
 Let us expand on this. Suppose one has a deformation of the Lorentz algebra Lie bracket $[X,Y]_{\kappa}$,
  parametrized by one continuous parameter $\kappa$, such that at $\kappa=0$ we recover the usual bracket.
   By taking a derivative of this bracket evaluated at  $\kappa=0$ one obtains an antisymmetric map $\psi(X,Y)\equiv\partial_{\kappa}[X,Y]_{\kappa=0}$. From Jacobi identity this map needs to be closed, so it satisfies the equation $\delta\psi(X,Y,Z)=0$, where
   \be
   \delta\psi(X,Y,Z)= \psi(X, [Y,Z]) +[X,\psi(Y,Z)] +\mathrm{cycl.}
   \ee
 where we should include a sum of cyclic permutations.
 There are obvious solutions of the closure equation given by infinitesimal redefinition of the generators, expressed as coboundaries: 
 \be
 \psi(X,Y)\equiv \delta\phi(X,Y)= [\phi(X),Y]+[X,\phi(Y)] - \phi([X,Y]).
 \ee
For a simple Lie algebra, the Chevalley-Eilenberg theorem states that the cohomology group is trivial, 
so we have that all closed forms $\psi$ labelling deformations come from a redefinition of the Lie algebra generators.
At the level of the group this essentially means that any deformed Lorentz group action on a manifold $\mathcal{P}$ is equivalent to 
the usual group action of Lorentz on $\mathcal{P}$, that is 
a map  $\mathrm{SO}(1,D)\to \mathrm{Diff}(\mathcal{P})$, which is a homeomorphism. In the following we restrict our discussion to $D=3$ for definiteness, although most of our results extend naturally to any  dimension.

We are interested in constructing a momentum space manifold $\mathcal{P}$ equipped with a metric $g$ and with an addition rule $\oplus: \mathcal{P}\times \mathcal{P} \to \mathcal{P}$, that also carries an action of the 
Lorentz group $p \to \Lambda(p)$, for $\Lambda \in \mathrm{SO}(1,D)$.
 This means is that there exists a mapping $\mathrm{SO}(1,D)\to \mathrm{Diff}(\mathcal{P})$
 that allows the Lorentz group to act by diffeomorphisms on $\mathcal{P}$.
 
 In order to construct Lorentz invariant theories, we are interested in group action that preserves both the metric and the addition rule. This is expressed by the conditions: \be\label{const}
 \Lambda^{*}(g)_{p}(X,Y)
 = g_{p}(X,Y),\qquad 
 \Lambda(p\oplus q) =\Lambda(p)\oplus \Lambda(q),
 \ee 
 where $X,Y\in T_{p}P$ are tangent vectors at $p$, and $\Lambda^{*}(g)_{p}(X,Y)=  g_{\Lambda(p) }(\rd_{p}\Lambda(X),\rd_{p}\Lambda(Y))$.
 The first condition expresses that  $\mathcal{P}$ can be locally written as product of $\mathbb{R}$ times a homogeneous $\mathrm{SO}(1,3)$ space, and the metric has locally the form of either $ \rd s^{2} = -\rd \tau^{2} + f^{2}(\tau) \rd \Omega(x)$, 
 or $ \rd s^{2} = \rd \lambda^{2} + f^{2}(\lambda) \rd \Omega(x)$, where $x$ are coordinates on the spacelike or timelike hyperboloid and $f$ is arbitrary.
 
 The second condition expresses the Lorentz invariance of the addition of momenta. We demand that Lorentz transformation is a morphism with respect to the non-linear addition. In physical terms the addition is the same before or after a boost.
 Note that this condition excludes the so-called $\kappa$-Minkowski spacetime \cite{Gubitosi:2011ej}.
 It is often stated that $\kappa$-Minkowski preserves Lorentz invariance. This is however not  true, in the sense that the 
 Lorentz transformation of the addition of two momenta {\it is not }   the addition of Lorentz transformed vector, in  $\kappa$-Minkowski.
 Instead, it satisfies $\Lambda(p \oplus_{\kappa} q) = \Lambda(p) \oplus_{\kappa} {T}_{p,\Lambda}\Lambda(q)$, 
 where $T_{p,\Lambda}$ is a transformation that depends on $p$ and $\oplus_{\kappa}$ denotes the $\kappa$-Minkowski addition.
 The property that the $\kappa$-Minkowski  addition satisfies is associativity.
 So it seems that the tension is between choosing a momentum addition which is associative but breaks Lorentz symmetry or a momentum addition which is non-associative but preserves Lorentz symmetry.
 We investigate the second option in this paper.
 
 The compatibility condition with respect to the addition can be conveniently written in terms of the left multiplication as 
 \be
\Lambda L_{p} \Lambda^{-1}= L_{\Lambda(p)}. 
 \ee
 If we denote by $I$ the identity of $\oplus$ and assume that there is only one identity,
 we see that this compatibility condition implies that the Lorentz action preserves it: $\Lambda(I)=I$.
 By considering the product of all possible left translations and their inverses, we can construct the group of left translations
 \be
 {\cal L} \equiv\{L_{p_{1}}^{\pm1}\cdots L_{p_{n}}^{\pm1}|\, p_{i}\in P \}.
 \ee
 We can also consider the subgroup of left translations which leave the identity invariant ${\cal G} \equiv \{ L \in {\cal L}|\, L(I)=I\}$. This group is left invariant by the adjoint action of the Lorentz group $\Lambda {\cal G} \Lambda^{-1}= {\cal G}$.
 
 It will be interesting to classify the most general solution to the set of constraints (\ref{const}). We reserve this for future work.
 In this paper we  study the simplest  solution to the set of constraints (\ref{const}). 
 We demand our particular solution to be homogeneous in the sense that  $ {\cal L}$ preserves the metric:
 \be\label{comp}
L_{p}^{*}(g) = g,
 \ee
   and we also demand  both ${\cal G}$ and $ {\cal L}$ to be finite dimensional groups given  by
 $
 {\cal G} = \mathrm{SO}(1,3)$, ${\cal L}= \mathrm{SO}(1,4).
 $
 Hence $\mathcal{P} = \mathrm{SO}(1,4)/\mathrm{SO}(1,3)$ is  de Sitter space.
In what follows, we will work in embedding coordinates, that is, we will consider de Sitter as a hyperboloid embedded in 5d Minkowski space. If $p$ is a point on de Sitter, then the Minkowski coordinates $P_A(p)$, $A=0,\ldots,4$ are constrained by
\begin{equation}
\eta^{AB} P_A P_B = \kappa^2,
\end{equation} 
where $\kappa$ is the radius of curvature. For simplicity we will work in units of $\kappa=1$.


\subsection{Addition rule}
Let us now focus on the  most general Lorentz covariant addition rule on de Sitter momentum space, which reduces to the addition in Minkowski space in the appropriate limit. The action of the addition will be implemented as an invariant subgroup of the manifold. Let us work in linear coordinates where the scalar product is denoted $\dd$ and corresponds to simply contraction with $\eta^{AB}$. We are now seeking an addition rule $P\oplus Q$ with left addition given by 
$ L_P\left(Q\right)=P\oplus Q $.
We chose the identity to be given by $I_A = \delta^4_A$, i.e. $L_P\left(I\right) = P$.
 In order to arrive at an addition rule that preserves the Lorentz invariance of the theory, we have to utilize all of the symmetries of the momentum space. For this reason, we have to require the addition to be metric compatible to achieve maximum symmetry. The condition (\ref{comp}) translates for the addition rule to the requirement that
\begin{equation}
L_P(Q)\cdot L_P(K) = Q\cdot K,
\end{equation}
which has to hold for all $Q$ and $K$. 
The Lorentz group is identified with the subgroup of translations preserving the identity $I$. This condition together with requirement that $L_P\in \mathrm{SO}(1,4)$ implies that $L_P$ has to be a tensor expressed only in terms of $P_A$ and $I_A$:
\begin{equation}
(L_P)^B_A = \delta^B_A + aP_A P^B + bI_A I^B + cP_A I^B + dI_A P^B,
\end{equation}
where $a, b, c$ and $d$ must be functions of the invariants $P\cdot P$, $I\cdot I$ and $P\cdot I$. However, since we have $P\cdot P = 1$ for any momentum $P_A$, effectively these parameters are just a function of $P\cdot I = P_4$.
We will now find the unique set of functions $a, b, c$ and $d$ that lead to an addition rule, which reduces to the addition on Minkowski space in the flat limit.

We have to impose that the addition does indeed treat $I$ as an identity. Solving the equation $L_P(I) = P$ gives us the conditions 
\begin{equation}
a(P\cdot I) + c =1, \ \ \ d(P\cdot I) + b = -1.
\end{equation}
The other property that we have to satisfy is the metric compatibility. Demanding $L_P$ to belong to O(4,1) gives us  three conditions
\begin{equation}
\begin{split}
a^2 + c^2 + 2a + 2a c\left(P\cdot I\right) =0,\\
b^2 + d^2 + 2 b +2b d\left(P\cdot I\right) =0, \\
\left(ab+cd\right)\left(P\cdot I\right) + \left(ad + bc\right) + c + d =0.
\end{split}
\end{equation}
 It is easy to show that these constraints have only two solutions: the first of them is $b+d=a+c=0$ and it gives us
\begin{equation}
(S_P)^B_A  =\delta^B_A +  \frac{(P-I)_A(P-I)^B}{ P\cdot I-1}.
\end{equation}
This solution turns out to be an inversion between $I$ and $P$, since a simple calculation shows that $\textnormal{det} S_P = -1$ and $S_P^2 = 1$, i.e. we have
$S_P\left(P\right) =I$. This solution is not physically interesting to us, as it does not reduce to the Minkowski addition in the 
limit where momenta are close to the identity. In particular, we see that  $P$ is its own inverse.

The other solution is the one of physical interest,  it reduces to addition on Minkowski space in flat limit, 
and is given by $b=d$ and $c=a+2$:
\begin{equation}
(L_P)^B_A  = \delta^B_A + 2 P_A I^B - \frac{(P+I)_A(P+I)^B}{1+ P\cdot I}.
\end{equation}
We can readily check that $L_P(I) = P$ is satisfied and that the determinant is positive in this case. 
\begin{equation}
P\oplus Q \equiv L_P(Q).
\end{equation}
Explicitly writing this out, we get a non-associative and non-commutative addition rule\footnote{Interestingly, we can relate the two additions: denoting by T the involutive operation $T\left(P_\mu,P_4\right) = \left(P_\mu,-P_4\right)$. We have
$
S_{P}= T L_{T(P)}.
$
}
\begin{equation}
\begin{split}
(P\oplus Q)_4 &= 2P_4 Q_4 - P\cdot Q, \\
(P\oplus Q)_\mu &= Q_\mu + P_\mu\frac{Q_4 + 2P_4 Q_4 - P\cdot Q}{1+P_4}.\label{eq:dSaddition}
\end{split}
\end{equation}
It can be easily checked that the inverse $\ominus$ of this addition rule is $\ominus P = (-P_\mu,P_4)$. A noteworthy property of this addition rule is that the inverse is a {\it morphism } for the addition:
\begin{equation}
\ominus (P\oplus Q) = (\ominus P)\oplus (\ominus Q).\label{eq:inversionprop}
\end{equation} 
This is a key difference from the  case of group manifolds (an example of which is $\kappa$-Poincar\'e), where $\ominus$ is an anti-morphism -- in the case of groups we have $\ominus (P\oplus Q) = (\ominus Q)\oplus (\ominus P)$.

A direct calculation also shows us that this rule is left invertible, that is it satisfies $\ominus P\oplus(P\oplus Q) = Q$, or equivalently $ L_{P}^{-1}= L_{\ominus P}$.
It is important to note that this is {\it not} true for the right multiplication.
 This can be easily seen from the fact that requiring both right and left invertibility necessarily leads to the inversion being an antimorphism as in a group manifold, not to eq. (\ref{eq:inversionprop}).

In order to analyse further this non-linear addition rule, let us first write it in terms of the 4-dimensional scalar 
product $P_{\mu}Q^{\mu}$, instead of the $5$ dimensional one $P\dd Q= P_{4}Q_{4}+ P_{\mu}Q^{\mu}$:
\bea
(P\oplus Q)_4 &=& P_4 Q_4 - P_\mu Q^\mu, \nn \\
(P\oplus Q)_\mu &=& Q_\mu+ P_\mu    - P_\mu\left(\frac{Q_\nu}{1+Q_4} + \frac{P_\nu }{1+P_4}\right)Q^\nu.\nn
\eea
For a classical particle, the momenta are timelike: $P_\mu P^\mu \equiv -P_{0}^{2} + P_{i}^{2} \leq 0$.
We have in this case 4 different sectors: first we can have positive or negative energy particle according to the sign of $P_{0}$, but we can also have positive or negative $P_{4}$.
 For such momenta we define the mass of a momentum  de Sitter particle to be given by\footnote{ By $m$ we mean $m/\kappa$ where $\kappa$ is a fundamental mass scale.} 
\be
P_\mu P^\mu = - \sinh^2 m,\qquad P_{4}=\pm\cosh m
\ee
The sector where $P_{4}<-1$ is a new sector that has no low energy analog.
In particular, as we are going to see there are  no geodesics connecting $P_4 \leq -1$ to the identity.
Let us finally note that since $P_4^2=1 - P_\mu P^\mu$, it is necessary for space like momenta to satisfy 
 $P_\mu P^\mu < 1$. Hence there is a natural cutoff of off-shell momenta.

In order to get a better understanding of this non-linear  addition let us focus on the physical sector where $P_{0}>0$ and $P_{4}>0$.
This sector possesses the property of being closed under addition, that is both $(P\oplus Q)_{4}$ and $(P\oplus Q)_{0}$ are positive if 
$P,Q$ are in the physical sector. Given two physical particles $P,Q$ of respectively mass $m,m'$,
 we define the centre of mass rapidity $\theta$ to be given by
$ P_{\mu}Q^{\mu}\equiv - \sinh m \sinh m' \cosh \theta. $ $\theta=0$ when the two particles are at rest with respect to each other.
In general $\theta$ represents the rapidity of the boost needed to go from the frame for which $P$ is at rest to the frame where $Q$ is at rest.

In terms of these data we have 
\be
(P\oplus Q)_4 =  \cosh m \cosh m' + \sinh m \sinh m' \cosh\theta.
\ee
Clearly when  the two particles are at rest with each other $(P\oplus Q)_4 = \cosh\left(m + m'\right)$ and  the total mass is simply the sum of the two masses.
This justifies a posteriori our definition of mass as $P_{4}=\cosh m$.
It also agrees with the Relative Locality definition of mass as the distance of $P$ from the origin.
\begin{figure}[ht]
	\centering
		\includegraphics[width=0.22\textwidth]{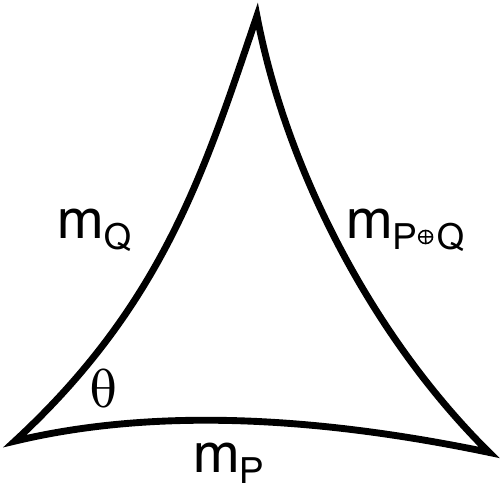}
	\caption{Mass addition in de Sitter for timelike momenta. Lengths of the edges are geodesic distances, which in Relative Locality are identified with masses of particles.}
	\label{fig:triangle}
\end{figure}
This can be visualised as a cosine law on de Sitter, see Figure \ref{fig:triangle}.

Let us for simplicity assume that both particles have the same mass $m=m'$, in this case 
the addition reads 
\bea
(P\oplus Q)_4 &=& 1 + 2 \sinh^{2}m \cosh^{2}\frac{\theta}{2}, \nn \\
(P\oplus Q)_\mu &=&  Q_\mu   + P_\mu \left(1 + \left(2 \sinh\frac{m}2 \cosh\frac\theta{2}\right)^{2}\right).
\eea
This expression makes it clear that in the centre of mass frame, where $(P\oplus Q)_i=0$, we do not have that $P_{i}=-Q_{i}$.
This makes for an intriguing property that follows from the non-commutativity of the addition.
It is also clear that if $m=0$ this reduces to the usual addition.

Let us now consider an expansion of the addition for small momenta. We can reinstate a mass scale $\kappa$ by 
defining $P_{\mu}=p_{\mu}/\kappa$, where $p$ is dimensionful,
considering the rescaled de Sitter space $P_{4}^{2} = 1 - P_{\mu}P^{\mu}/ \kappa^2$.
Expanding the addition  in $\kappa^{-1}$ we get 
\begin{equation}
\begin{split}
(P\oplus Q)_\mu &= q_\mu + p_\mu + \frac1{2 \kappa^{2}} p_{\mu} (q_{\nu}+p_{\nu})q^{\nu}+ 
 \frac1{8 \kappa^{4}} p_{\mu} (q_{\nu}q^{2}+p_{\nu}p^{2})q^{\nu} +\cdots \end{split}
\end{equation}
This addition clearly reduces to  the addition rule in a flat momentum space, in the low energy limit.
We have thus constructed the unique  Lorentz covariant addition rule on de Sitter space, 
compatible with the metric  that reduces to the addition on Minkowski space in the low energy  limit.

In many applications it is useful to know the addition of collinear vectors.
We assume that $P_{\mu}=\sinh a n_{\mu}$ and $ Q_{\mu}=\sinh b n_{\mu}$, where $n_{\mu}n^{\mu}=-1 $ is a unit 
4-vector. We can readily see that 
$$(\sinh a n_{\mu} \oplus \sinh b n_{\mu}) = \sinh(a+b) n_{\mu}.$$ 
In particular, if we sum $n$ times the same vector $p =\sinh a n$ we obtain that 
\be
\underbrace{(p\oplus (p\oplus \cdots \oplus p)\cdots ))}_{n}{}_{\mu} = \sinh(n a) n_{\mu}.
\ee
If one works with normal coordinates $\tilde{p}_{\mu}= \sinh^{-1} |p| \frac{p_{\mu}}{|p|}$, the collinear addition simply becomes the usual addition: $\underbrace{(\tilde{p}\oplus (\tilde{p}\oplus \cdots \oplus \tilde{p})\cdots ))}_{n}= n\tilde{p}$.

It is clear from the definition that the addition is not commutative. What is less clear but also true is that this addition is also non associative. The non associativity will be related to the curvature of de Sitter.

\subsection{Geometric understanding of the addition}
\begin{figure}[h]
	\centering
		\includegraphics[width=0.20\textwidth]{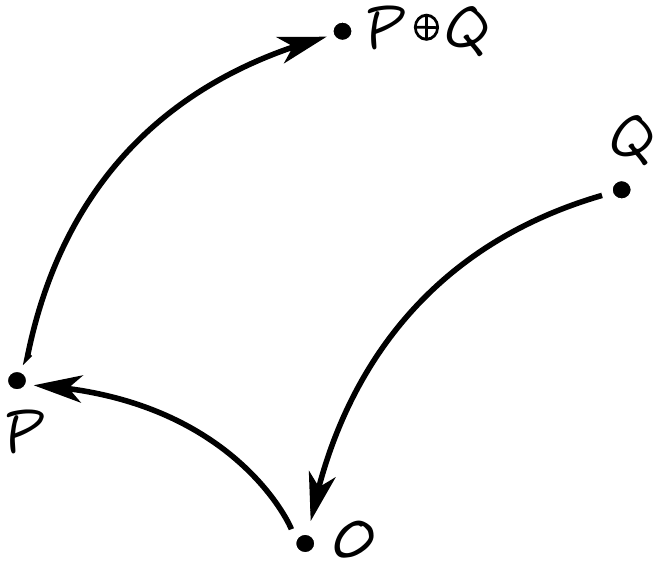}
	\caption{The addition rule seen as parallel transports along geodesics.}
		\label{fig:addition}
\end{figure}

\begin{figure}[ht]
	\centering
		\includegraphics[width=0.50\textwidth]{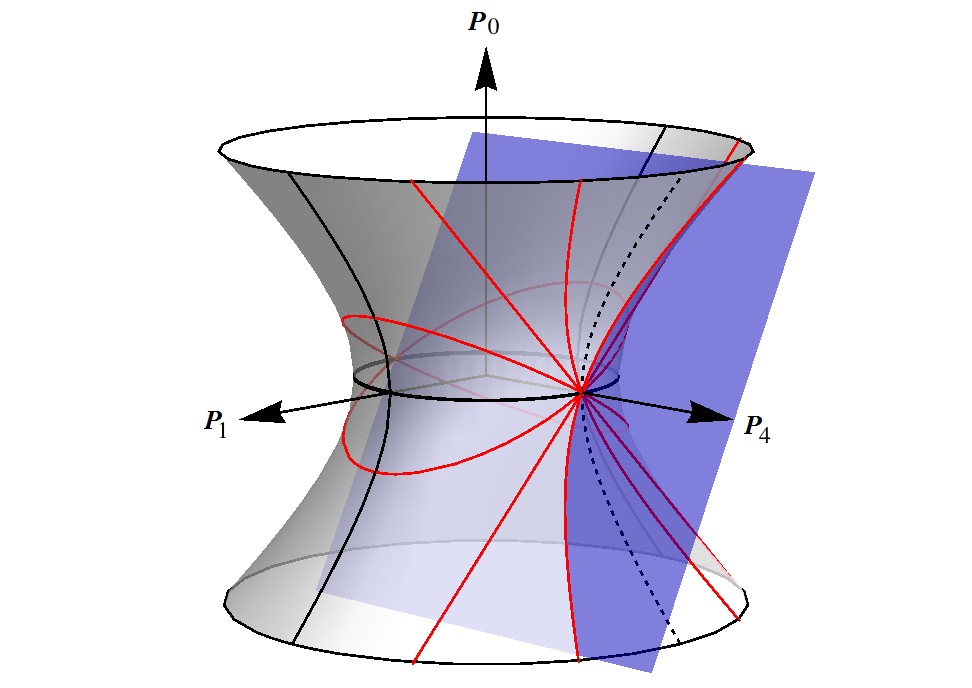}
	\caption{Geodesics on de Sitter in red as the intersections of the hyperboloid with planes passing through the $P_4$ axis.
	Figure taken from \cite{Gubitosi:2011ej}.}
		\label{fig:dSgeodesics}
\end{figure}

In this section we will study the geometrical interpretation of the addition rule  (\ref{eq:dSaddition}). We will prove that the addition $P\oplus Q$ is the geodesic addition.
It can be obtained by thinking of $P$ and $Q$ as geodesics from the origin, and constructing $P\oplus Q$ as the 
geodesic starting from $P$ parallel to the $Q$ geodesic.
 This can be seen in Figure \ref{fig:addition} and is given  by
\begin{equation}
P\oplus Q \equiv \exp_P \circ \  U^I_P \circ \exp_I^{-1} Q. \label{eq:exponentialaddition}
\end{equation}
where $\exp_{a}Q$ is the exponential map at $a$, and $U^{I}_{P}$ is the parallel transport operator from the origin $I$ to $P$.
This expression can be thought of as first representing  $Q$ as the tangent vector to the $Q$ geodesic at  the origin, 
then parallel transporting  it to $P$, and finally using this vector to construct a  geodesic starting from $P$
which ends at $P\oplus Q$.
Before we go on to prove this expression, we have to set up some facts about the geodesics in de Sitter space. There is a nice geometrical view of the geodesics in this space if we consider dS as embedded in the 5-dimensional Minkowski space.
The geodesics from origin of the embedding coordinates $I_A = \delta_A^4$ are given by intersections of the hyperboloid with planes through the $P_4$ axis, as can be seen in Fig. \ref{fig:dSgeodesics}. 
In order to find the  geodesic equation for de Sitter space, we simply consider the Lagrangian for the goedesic
\begin{equation}
L = \frac{1}{2}\dot{P}^2 + \frac\Lambda2\left(P^2 - 1\right),
\end{equation}
where $\Lambda$ is a Lagrange multiplier constraining the geodesic to stay on the hyperboloid. We easily see that the Euler-Lagrange equation (the geodesic equation) is
\begin{equation}
\ddot{P}_A = \Lambda P_A.
\end{equation}
In order to identify  $\Lambda$ let us notice that the following ``angular momentum'' tensor of the geodesic is  conserved:
\begin{equation}
J_{AB} \equiv \dot{P}_A P_B - P_A\dot{P}_B.
\end{equation}
It is easy to check that this tensor obeys the following relations
\begin{equation}
J_{AB}P^B = \dot{P}_A, \ \ \ J_{AB}\dot{P}^B = -\dot{P}^2 P_A,
\end{equation}
which together give us the geodesic equation in a very simple form
\begin{equation}
\ddot{P}_A + \dot{P}^2 P_A = 0.\label{eq:geodesiceqn}
\end{equation}
We have thus found the value of $\Lambda$ to be $\dot{P}^2 $. The Euler-Lagrange equations of motion for the Lagrange multiplier imply however that   $\dot{P}^2 = const$. Since we also have that $P\dd\dot{P}=0$, and $P^{2}=1$, 
this constant is negative for a timelike geodesic and positive for a spacelike geodesic.
 We will choose this constant to be $-1 $ in the timelike case and $+1$ in the spacelike case.
Solving (\ref{eq:geodesiceqn}) implies that timelike and spacelike geodesics starting at $P$ with velocity $\dot{P}$ are respectively given by
\be\label{geo}
P(\tau) = \dot{P}\sinh \tau + P \cosh \tau,\qquad P(\ell) = \dot{P}\sin \ell + P \cos \ell.
\ee
$\tau $ is the proper time along the geodesic between $P$ and $P(\tau)$ and $\ell$ the proper distance.
 It is characterized by $P\dd P(\tau)=\cosh\tau$, or $P\dd P(\ell)=\cos \ell$.
This shows that two points $P,Q$ can be joined by a timelike geodesics iff $P\dd Q \geq 1$,
they can be joined by a spacelike geodesic if  $|P\dd Q| \leq 1$ and they cannot be joined by a geodesic if $P\dd Q <-1$.
In summary  the  distance $d$ between $P$ and $Q$ is given by
\begin{equation}
\cosh{d} = P\dd Q, \qquad \cos{d} = P\dd Q, \label{eq:coshd}
\end{equation}
depending on whether they are timelike or spacelike separated.

 The exponential map  is a mapping $\exp_P : T_P \mathcal{M} \rightarrow \mathcal{M}$ that maps us to different points on the geodesic defined by the tangent vector $X\in T_P \mathcal{M} $. The tangent space is defined as $T_P\mathcal{M} = \{X | X \cdot P = 0 \}$. Given an element of $T_P\mathcal{M}$ we denote by $\tau_{X}$ its norm\footnote{If $X$ is spacelike, we take $\tau_{X}$ to be pure imaginary.}: $X^{2}\equiv - \tau_{X}^{2}$, and we denote $\hat{X}\equiv X/\tau_{X}$ the corresponding normalized element. The map (\ref{geo}) represents the exponential map:
 \be
 \exp_{P}(X) \equiv \hat{X} \sinh \tau_{X} + P \cosh \tau_{{X}}.
 \ee
 Let us finally note that we can characterise a geodesic by its starting and ending point instead of its initial velocity.
 Solving for $\exp_{P}X = Q$, we get 
 \be
\exp_{P}^{-1}(Q) = \frac{\tau}{\sinh \tau }\left(Q -\cosh \tau  P \right),\qquad \cosh \tau = P\dd Q.
 \ee
 The geodesic written in terms of this data is 
 \be\label{geod}
P(t) = \frac{1}{\sinh{\tau}}\left(\sinh{\left(\tau - t \right)}P + \sinh{t}Q \right)
 \ee
 A vector $V^{A}$  is said to be transported parallely along a geodesic if it satisfies $P\dd V=0$ and
 \be
\partial_{t}V^{A}+  P^{A} (\dot{P}\dd {V})=0.
 \ee
 This can be solved by $ V(t) = U_{P}^{P(t)}\dd  V(0)$, where $U_{P}^{P(t)}$ is the parallel transport operator from $P$ to $P(\tau)$ and is a solution of
 \be
\partial_{t}U_{P}^{P(t)} = J\dd U_{P}^{P(t)},
 \ee
 where $J_{AB}= \dot{P}_A P_B - P_A\dot{P}_B $  is the angular momentum tensor.
  The solution of the parallel transport equation is then $ U_{P}^{P(t)} = e^{t J}$, and since $J^{3}=J$ this can be easily evaluated as 
  \be
 ( U_{P}^{P(t)})^{A}{}_{B} = \delta^{A}_{B} + \sinh t (\dot{P}^{A} P_B - P^{A}\dot{P}_B) + (\cosh t -1)  ({P}^{A} P_B - \dot{P}^{A}\dot{P}_B).
  \ee
 Given  the expression (\ref{geod}) for a geodesic from $P$ to $Q=P(\tau)$, we can express $\dot{P}= (  Q - \cosh \tau P)/\sinh \tau$ with $ \cosh \tau =P\dd Q$, thus
 $J_{AB} = (Q_{A}P_{B}-P_{A}Q_{B})/\sinh\tau $ and  we obtain that
 \bea
 ( U_{P}^{Q})^{A}{}_{B} 
 &=& \delta^{A}_{B}+ 2 Q^{A} P_B - \frac{(P+Q)^{A}(P+Q)_{B}}{1+P\dd Q}.\label{eq:HTrans}
 \eea
 We recognize here that the left addition operation is essentially given by the parallel transport $U_{I}^{P}$.
 
 We can now compute the composition
 \bea
 \exp_{P}\circ U_{I}^{P}\circ \exp_{I}(Q) = \exp_{P}\circ U_{I}^{P}\dd \frac{(Q- \cosh \tau I)}{\sinh \tau}
 = \exp_{P} \left(\frac{((P\oplus Q)- \cosh \tau P )}{\sinh \tau}  \right)=P\oplus Q,
 \eea
as expected.
The last equality follows from the identity $ (P\oplus Q) \dd P = Q\dd I = \cosh \tau$, which reflects the fact that the parallel transport preserves the metric.
This gives us a geometrical basis for the composition rule, it is given by the geodesic composition that exists in any metrical manifold.

\section{Noncommutative spacetime}
In what follows we will try to find the spacetime structure emergent from the Relative Locality dynamics on a de Sitter momentum space with a metric compatible addition rule. This will give us a  first example of momentum space geometry in Relative Locality which preserves full action of Lorentz transformations and is non-trivial.
The goal of this section is to understand the structure of space time associated with curved momentum space.
The phase space structure is the cotangent structure $T^{*}\mathcal{P}$.

\subsection{Coordinates and frames on de Sitter}

We can describe points in de Sitter either in terms of the embedding coordinates $P_{A}$ which are 5-dimensional and constrained to satisfy $P_{A}P^{A}=1$ or in terms of local 4-dimensional coordinates $p_{\mu}$.
In the embedding framework a tangent vector $X \in T_{P}\cal{M}$ is given by a 5d vector orthogonal to $P$.
The relationship $P^{A}(p)$ between these two frameworks is encoded into frame fields 
$$e^\mu_A (p) \equiv \partial^\mu P_A. $$ This frame satisfies $ P^{A}e_{A}^{\mu}(p) =0$.

We can invert the relation between $p$ and $P$ using the the ``inverse frame''  $e^A_\mu (p)$,  defined to be the solution of \begin{equation}
e^\mu_A (p)e^A_\nu (p) = \delta^\mu_\nu, \ \ \qquad  e^A_\mu (p)e^\mu_B (p)  = \delta^A_B - P^A P_B. \ \ \ 
\label{eq:jacobians}
\end{equation}
These inverse frames also satisfy the constraints $ P_{A}e^{A}_{\mu}(p)=0$.
An explicit way to represent the inverse relation   $p_\mu(P)$ is to impose that $p^{\mu}$ is  a homogenous functions of the 5-d Minkowski coordinates. Explicitly this means that we require $p_\mu(\lambda P) = p_\mu(P)$ for any $\lambda > 0$.
The inverse frame is then given by $e^{A}_{\mu}(p) = \pa^{A}p_{\mu}$.

Using the frames we can find expressions for a vector in the  different pictures:
\begin{equation}
V^A(P) = e^A_\mu (p) V^\mu (p), \ \ \quad V^\mu (p) = e_A^\mu (p) V^A(P).
\end{equation}
Similarly, we can express tensors of other rank in different coordinates by applying the Jacobians, for example the metric and its 
inverse are given by
\begin{equation}
g^{\mu\nu} = e_A^\mu \eta^{AB} e_B^\nu, \ \ \ g_{\mu\nu} = e^A_\mu \eta_{AB} e^B_\nu .\label{eq:metric}
\end{equation}
Note that $e$'s allow us to compute the metric connection in a simple way
$
 \Gamma^{\alpha\beta}_\mu (p) = \left(\partial^\alpha e^\beta_A \right)e^A_\mu (p).\label{eq:connection}
$
Finally, and that is going to be the main use for us, we can use the frames to 
get  expression for the parallel transport operator from $0$ to $p$ in the 4d coordinates
\begin{equation}
(U^0_p)^\nu_\mu = e^A_\mu (p) (U_P^{I})^B_A e_B^\nu (0) 
= e_{\mu}^{(\nu)}(p) - \frac{e_{\mu}^{4}(p)  P^\nu}{1+ P_4}.\label{eq:paremb}
\end{equation}
This evaluation simplifies a lot thanks to the conditions $ P_{A}e^{A}_{\mu}(p)=0$ and $I^{B}e^{\mu}_{B}(0)=0$. Also,
 we assumed that we have coordinates for which $e_{A}^{\nu}(0)=\delta^{\nu}_{A} $.

In order to evaluate these expressions we need to chose a specific set of coordinates. Note that the action of Relative locality is invariant under deffeomorphisms in the momentum space, so the sense in which coordinates are physical is the same as the one in General Relativity.
The most general set of coordinates that {\it  preserves} the linearity of the Lorentz group action and map $I$ to the origin  is labelled by a function $\alpha(p^{2})$, where $p^{2}\equiv p_{\mu}p^{\nu}\eta^{\mu\nu}$ and given by 
$$ P_{\mu}= \alpha p_{\mu},\quad P_{4}=\sqrt{1-\alpha^{2}p^{2}}.$$
There are several natural choices for $\alpha$, the simplest, computation wise, is $\alpha=1$ which corresponds to taking the embedding coordinates.
In this coordinates the inverse metric is $g_{\mu\nu} =\eta_{\mu\nu} - p_{\mu}p_{\nu}$.
Another natural choice are the normal coordinates, where $\alpha = \sinh|p|/|p|$. These are coordinates for which the collinear addition is trivial. Finally the most important set of coordinates are the conformal coordinates, for which the metric is conformally flat
$g^{\mu\nu} = \alpha^{2} \eta^{\mu\nu}$.
The conformal coordinates correspond to a choice $\alpha = 2/(1+p^{2})$, that is 
\be
\alpha=1+P_{4},\qquad
P_{\mu} = \frac{2p_{\mu}}{1+p^{2}},\qquad P_{4}=\frac{1-p^{2}}{1+p^{2}}.
\ee
 
From the definition we can explicitly evaluate the frames to evaluate the parallel transport operator.
Let us first do that for the embedding coordinates $P_{\mu}=p_{\mu}$. In these coordinates, the frames and their inverses are given by:
\bea
e^\mu_{4} (p) &=& -\frac{p^\mu}{P_4}, \qquad  \,\,e^\mu_{(\nu)} (p) = \delta^\mu_\nu \\
	e_\mu^{4} (p) &=& -p_\mu P_4, \qquad  e_\mu^{(\nu)} (p) = \delta_\mu^\nu - p_\mu p^\nu. \label{eq:embeddingjacobians}
\eea
We can calculate, in these coordinates, the metric, its inverse and the parallel transport  to be 
\begin{equation}
g^{\mu\nu} = \eta^{\mu\nu} + \frac{p^\mu p^\nu}{P^2_4}, \quad g_{\mu\nu} = \eta_{\mu\nu} - p_{\mu}p_\nu,
\quad (U^0_p)^\nu_\mu 
= \delta_{\mu}^{\nu} - \frac{p_{\mu}  p^\nu}{1+ P_4}.
\end{equation}
Note that we raise indices with $\eta$: $p^\mu = \eta^{\mu\nu} p_\nu$.

In general Lorentz invariant coordinates $P_{\mu}=\alpha p_{\mu}$ the frame is related to the previous one by the transformations 
\be
e_{A}^{\mu}(p) \to \alpha \left[ \delta_{\nu}^{\mu}+ \frac{2\alpha'}{\alpha} p_{\nu}p^{\mu}\right]e_{A}^{\nu}(\alpha p),
\ee
We can use this to evaluate  the frame 
in conformal  coordinates. The expressions simplify to
\bea
e^{\mu}_{4} (p) &=& - (1+P_{4})^{2}{p^{\mu}}, \qquad  \,\,e_{(\beta)}^\mu (p) ={(1+P_{4})} \left[  {\delta_{\beta}^{\mu}}-(1+P_{4}) {p_{\beta}p^{\mu}}\right],
\eea
while the 
inverse frame field in conformal coordinates is
\bea
e_{\mu}^{4} (p) &=& - {p_{\mu}}, \qquad  \,\,e_\mu^{(\nu)} (p) =\frac1{(1+P_{4})} \left[  {\delta_{\mu}^{\nu}}-(1+P_{4}) {p_{\mu}p^{\nu}}\right].
\eea
Thus we conclude that the metric and parallel transport in conformal coordinates are simply given by
\be
g^{\mu\nu}= (1+P_{4})^{2}\eta^{\mu\nu},\qquad \quad (U^0_p)^\nu_\mu 
= \frac{\delta_{\mu}^{\nu} }{(1+P_{4})}.
\ee
The conformal coordinates are particularly simple, since the parallel transport is simply a rescaling.

Finally evaluating the curvature, we get:
\begin{equation}
	R^{\mu\alpha \nu \beta} = g^{\alpha\beta}g^{\nu\mu} - g^{\alpha\nu}g^{\beta\mu},\label{eq:riemann}
\end{equation}
as we would expect for a de Sitter space, a homogenous manifold.


\subsection{Interaction coordinates}

 Let us start by investigating the properties of the cotangent space $T^* \mathcal{P}$.
 At a given point $p$, the cotangent plane  $T^*_p \mathcal{P}$ is the configuration space.
  The  linear coordinates $x^\mu\in T^*_p \mathcal{P}$ on configuration space are conjugate to $p_\mu$:
  \be
  \{p_{\mu}, x^{\nu}\}=\delta _{\mu}^{\nu}.
  \ee
  Naively we would like to define space time to be the set of configuration variables, that is $T^{*}_{p}\cal P$.
  There are however two related issues with that.
  First, this definition depends on the value of $p$, so we cannot assign one space-time independently of the value of the momenta.
   A related issue is the fact that the coordinates $x^{\mu}$ are not invariant under momentum space diffeomorphisms.
   For instance, if we stay within Lorentz covariant coordinates and consider the transformation $p_{\mu}\to \bar{p}_{\mu}= \alpha(p^{2}) p_{\mu}$,
   the configuration space coordinates transform as 
   \be\label{cano}
   x^{\mu}\to \bar{x}^{\mu}=\frac1{\alpha}\left(x^{\mu} - \frac{2\alpha'}{(\alpha + 2\alpha' p^{2})} (p\dd x) p^{\mu}\right),
   \ee
  so that $\{\bar{p}_{\mu},\bar{x}^{\nu}\} = \delta_{\mu}^{\nu}.$
  This is not satisfactory, since it would mean that the notion of space-time is strongly dependent of the coordinate chosen in momentum space. Note that these diffeomorphisms act on $x$ in a way that depends on $p$.
  
\begin{figure}[ht]
	\centering
		\includegraphics[width=0.25\textwidth]{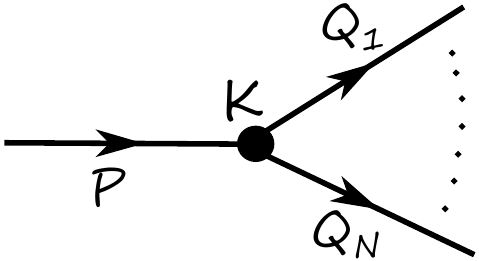}
	\caption{A subprocess of a tree of events leading to Snyder commutation relations for interaction coordinates. A particle of momentum $P$ decays to N particles of momenta $Q_1,\ldots Q_N$.}
		\label{fig:spacetime}
\end{figure}

 A  resolution of this puzzle was already proposed in \cite{AmelinoCamelia:2011bm}.  The idea is to define spacetime as the tangent space at the identity\footnote{Note that this implies that the base point of the tangential space is now special in some sense, possibly making this spacetime not homogenous.}: $M=T^{*}_{0}{\cal P}$.
  This corresponds to going to the interaction picture, since all interactions take place at $0$ momenta.
   The interaction coordinates $z$ are vector fields corresponding to coordinates transported from some cotangent plane to the origin of momentum space.  In order  to transport the configuration coordinates $x\in T^{*}_{p}\cal{P}$ to $M$ and assuming that we chose a parallel transport operation,  we need to chose a path that goes from $p$ to $0$.
   Indeed, since our momentum space is curved, the transport  is  path dependent. 
   A natural choice is to parallel transport along a geodesic from $p$ to origin, which is given by our transport operator $U_p^{0}$.
   We can therefore define the interaction coordinate to be
   \be\label{zco}
   z^{\nu}\equiv  x^{\mu} (U_{p}^{0})_{\mu}^{\nu}.
   \ee
 In Relative Locality such an operator should come from considering some conservation law, or more generally, a specific process. In general, with different conservation laws we can arrive at different rules of transport from $p$ to $0$. 
 We find that if (and only if) we build  a process for which the conservation laws take  the specific form $\mathcal{K} = \ominus\left(Q_1\oplus(\cdots\oplus Q_N\right)) \oplus P$, do we get an operator which acts as parallel transport from $P$ to 0. In this case the interaction coordinates are given by the previous relation. Such a conservation law is shown in Figure \ref{fig:spacetime}. 
The advantage of the interaction coordinates is that  under momentum space diffeomorphisms preserving the identity, they transform linearly and homogeneously $z\to \Lambda z $, where $\Lambda$ is a constant matrix representing the differential of the diffeomorphism at $0$.

If we denote by $p_{\mu}$ the embedding coordinates and $\bar{p}_{\mu}$ the conformal coordinates, and $x^{\mu},\bar x^{\mu}$ the corresponding conjugate variables, we can evaluate the interaction coordinate to be given by
\be
z^\nu  = (1+\bar{p}^{2})\frac{\bar{x}^{\nu}}{2},\quad \mathrm{or}\quad z^{\nu}= x^\nu - \frac{(x\dd p) p^\nu}{1+P_{4}}.\label{eq:z}
\ee
We can check that the two expressions agree  according to law  of canonical transformation (\ref{cano}),
since $\bar{p}= p/(1+P_{4})$.
We can evaluate the Poisson bracket of the interaction coordinates and we find it to be
\be
\{z^{\mu},z^{\nu}\} = \frac{J^{\mu\nu}}{1+P_{4}},\qquad J^{\mu\nu}\equiv (x^{\mu}p^{\nu}-x^{\nu}p^{\mu})=(\bar{x}^{\mu}\bar{p}^{\nu}-\bar{x}^{\nu}\bar{p}^{\mu}). \label{eq:zSnyderBracket}
\ee
This Poisson bracket is reminiscent of Snyder commutation relations.
Interestingly enough, it is not identical to Snyder, since there is an extra factor $1/(1+P_{4})$ multiplying the angular momenta.
This extra factor comes from the parallel transport to the origin.
This shows that Snyder model, which is not derived from any fundamental principle, may be too naive.  For details on Snyder's quantized spacetime, see the Appendix.


\subsection{Embeding noncommutativity}

It is now interesting, for completeness, to express our results from the 5-dimensional point of view (see \cite{FGL} for a related discussion) .
We have seen  that the cotangent planes $T^*_p \mathcal{P}$ can be described as $T_{P}^{*}{\cal P}=\{ X^{A}| X\dd P =0\}$, with
the relation between the vector fields $x^{\mu}$ and $X^{A}$  given by (using the embedding coordinates)
\begin{equation}
x^\mu = e^\mu_A X^A = X^\mu - \frac{X^4}{P_4}P^\mu.
\end{equation}
The inverse relation of the 5d coordinates as functions of the 4d coordinates is $X^{A}= e^{A}_{\mu} x^{\mu}$, which gives explicitly: 
\begin{equation}
X^\mu = x^\mu - (x^\nu p_\nu)p^\mu, \qquad X^4 = - (x^\nu p_\nu)P_4.
\end{equation}
Since $x^\mu$ and $p_\mu$ are canonically conjugate, we find that the 5d vector fields do not commute and satisfy
\begin{equation}
\left\{P_{A},X^{B}\right\} = \delta_{A}^{B} - P_A P^B, \ \ \ \left\{X^A,X^B\right\} = J^{AB},\label{eq:XPbrackets}
\end{equation}
where $J^{AB} \equiv X^A P^B - X^B P^A $ are the generators of ten dimensional de Sitter algebra. 
These Poisson brackets are consistent with the relations $ X^{A}= J^{AB}P_{B}$.
 This looks exactly like the  structure of commutators of Snyder spacetime \cite{Snyder:1946qz}, 
 which was the first ever quantized spacetime to be studied. 
 The price to pay however is that again  $X^{A}$, defined as a vector field,
  depends on the momenta $P$, since it is always orthogonal to $P$.
 
 The cure, as we know, is to transport back the vector field to the origin and work with interactions co-ordinates defined by 
 \be
 Z^B = X^A U_A^B\quad \to\quad Z^\mu = X^\mu - \frac{ X^4 P^\mu}{1+P_4},\qquad Z^{4}=0.
 \ee
 As expected these interaction coordinates live in a fixed plane $Z_{4}=0$, which can serve as a definition for space-time.
 We observe that these interaction coordinates satisfy a modification of Snyder algebra which involves the factor $1/(1+P_{4})$.
 That is:
 \be
 \{Z^\mu,P_{\nu}\}=\delta^{\mu}_{\nu}- \frac{ P^\mu P_{\nu}}{1+P_4}, \qquad \{Z^\mu,Z^{\nu}\}= \frac{J^{\mu\nu}}{1+P_{4}}, \qquad \{Z^4,Z^A\} = \{Z^4,P_A\}=0.
 \ee
 This shows that the extra factor is necessary in order to go to a well defined, space-time picture which is $P$ independent.


\subsection{Process interactions}

\begin{figure}[ht]
	\centering
		\includegraphics[width=0.25\textwidth]{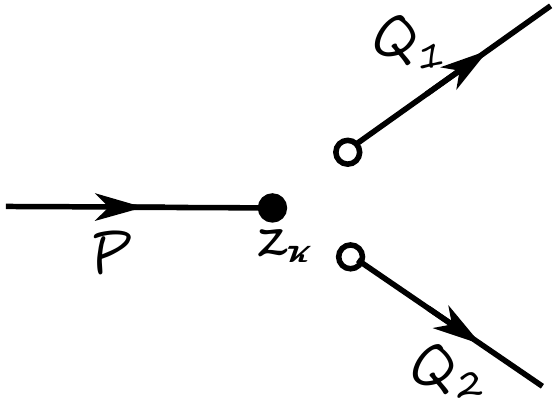}
	\caption{When we try to embed a process in a single spacetime, we end up with a nonlocal vertex, in which two of the particles' worldlines are not connected to the event at which they are created.}
		\label{fig:nonlocal}
\end{figure}

We now study more closely the transports associated with an interaction vertex. 
We focus on the interaction of 3 particles with 2 incoming momenta $Q_{1},Q_{2}$ and one outgoing momentum $P$, with
 a trivalent vertex with total conservation rule given by $\mathcal{K} = \ominus\left(Q_1 \oplus Q_2\right)\oplus P$.
 This vertex defines an event connecting different particles' worldlines.
 We assign to each worldline a configuration variable $x_{P}\in T_{P}^{*}{\cal P}$, $x_{Q_{i}}\in T^{*}_{Q_{i}}{\cal P}$ 
 which label the end points of the worldlines.
 These different configuration coordinates belong to different cotangent planes and need to be related to the interaction coordinate $z$. Acoording to Eq. (\ref{eq:relxz}), the prescription for going between interaction and worldline coordinates is given by:
\begin{equation}
\begin{split}
x_P &= z_{\mathcal{K}} \frac{\partial\mathcal{K}}{\partial p} = z U_0^p \\ 
x_{Q_1} &= -z \frac{\partial\mathcal{K}}{\partial q_1} = z U_0^p V^{q_1}_p \\
x_{Q_2} &= - z  \frac{\partial\mathcal{K}}{\partial q_2} = z U_0^p U^{q_2}_p, \label{eq:3vertexevent}
\end{split}
\end{equation}
where for the last two we used the property of Eq. (\ref{eq:UVrelation}). 
Note that it is only for the  outgoing momenta that we have the relation $z= x_P U_p^0$ studied in the previous section.
In order to understand more the structure of the vertex, let us write the relations in the interaction plane, defined by 
$z_{P}\equiv x_P U_p^0$, $z_{Q_{i}}\equiv x_{Q_{i}}U_{q_{i}}^{0}$. 
We see that even if we map the relations to the interaction plane, not all vertices agree. Indeed, defining 
$H_{1}\equiv U_0^p V^{q_1}_pU^{q_{1}}_{0}$ and $H_{2}\equiv U_0^p U^{q_2}_p U^{q_{2}}_{0}$,  which are operators acting on
the interaction plane $T^{*}_{0}{\cal P}$, we get
\be
z_{P}=z,\qquad  z_{Q_{i}}= zH_{i}.
\ee
The operators $H_{i}$ represent holonomies in momentum space from the origin to the origin.
This  means that the different worldlines, when mapped to the interaction plane by the same transport operator, do not 
overlap.  Thus we get a nonlocal vertex in the interaction plane, as seen in Figure \ref{fig:nonlocal}. The only case in which the particles do meet is if $z=0$.
The relativity of locality tells us that there is no universal spacetime. If we want to map all of the physics to one plane, the spacetime that emerges is noncommutative and nonlocal.  At the quantum level the coordinates would acquire an uncertainy relation and it is an interesting possibility to investigate whether this non locality can be hidden behind 
quantum uncertainty. 
To better understand this nonlocality, we have to investigate the properties of the different transport operators, which takes us to the next section.


\section{Building blocks for calculations}
In this section we will build the machinery for Relative Locality calculation on our de Sitter momentum space. The main objects used in such calculations are the transport operators defined in Eq. (\ref{eq:lefttr}) and Eq. (\ref{eq:righttr}), and the operators related to them. The nontrivial property of the addition rule (\ref{eq:dSaddition}) is its left-but not right invertibility. This means that there will be qualitative difference between transporting from right or left.

\subsection{Transport operators}
We will now proceed to calculate the transport operators as defined by Eq. (\ref{eq:lefttr}) and Eq. (\ref{eq:righttr}). We will work here in the 4d coordinates defined in previous section by $P_\mu = p_\mu$. By taking the addition rule written in these coordinates and differentiating $(p\oplus q)$ with respect to $q$ or $p$,  we get the following operators easily

\begin{equation}
\begin{split}
\left(U^q_{p\oplus q}\right)^\nu_\mu &= \delta^\nu_\mu - \frac{p_\mu p^\nu}{1+P_4} - \frac{p_\mu q^\nu}{Q_4}, \\
\left(V^p_{p\oplus q}\right)^\nu_\mu  & = Q_{4}\left( \delta_{\mu}^{\nu}- \frac{p_{\mu}q^{\nu}}{(1+P_{4})Q_{4}}\right)
- \frac{p\dd q}{(1+P_{4})}\left( \delta_{\mu}^{\nu}+ \frac{p_{\mu}p^{\nu}}{(1+P_{4})Q_{4}}\right).
\label{eq:leftright}
\end{split}
\end{equation}
where we denote  $p \dd q \equiv p_{\alpha}q^{\alpha}$, not to be confused with the similar notation of the $5d$ scalar product.
By setting either $p,q$ or $p\oplus q$ to 0 in this equation, we easily get left and right transports to and from the origin
\begin{align}
\left(U^p_{0}\right)^\nu_\mu &=   \delta^\nu_\mu + \frac{p_\mu p^\nu}{P_{4}(1+P_4)},\qquad \quad
\left(U^0_{p}\right)^\nu_\mu = \delta^\nu_\mu - \frac{p_\mu p^\nu}{(1+P_4)},\nn \\
\left(V^p_{0}\right)^\nu_\mu &=   \delta^\nu_\mu + \frac{p_\mu p^\nu}{P_4(1+P_4)},\qquad\quad
\left(V^0_{p}\right)^\nu_\mu = P_4 \delta^\nu_\mu \label{eq:strange}. 
\end{align}
We can easily see that $ U_{p}^{0}$ and $U_{0}^{p}$ are inverses of each other, which follows from the left invertibility property.
However this is no longer true for $V_0^p$ and $V_p^0$, which is a consequence of the lack of right invertibility of the addition rule. We also see that $ U$ is metrical, that is $[(U_{p}^{0})^{T}\eta(U_{p}^{0})]_{\mu\nu}= \eta_{\mu\nu}-p_{\mu}p_{ \nu}= g_{\mu\nu}(p)$. This is true for $V^{p}_{0}$, but no longer true for $V_{p}^{0}$.
Another interesting property is that transport between 0 and $p$ is the same as between 0 and $\ominus p$. This follows from the fact that in our coordinates $\ominus p = -p$.

In order to construct the   transports between a point $p$ and point $q$ we need to invert both the left and right addition.
We start with the equation  $(p\oplus q) = r$ and solve for both $p= R_{q}^{-1}(r)$ and $q=L_{p}^{-1}(r)$.
We find 
\begin{equation}
\begin{split}
q_\mu &= r_\mu - \left(r_4 + \frac{p\!\cdot\! r}{1+p_4}\right)p_\mu =  L_{p}^{-1}(r) \\
p_\mu &= \left(r_\mu - q_\mu\right)\frac{q_4+r_4}{1+q_4 r_4 - q\!\cdot\! r} = R_{q}^{-1}(r)\label{eq:invert}
\end{split}
\end{equation}
Checking the validity of first one is trivial as it is just $q = \ominus p \oplus r$,  the second,  can be checked directly.
It is useful to note that we can get $p_4$  from 
\begin{equation}
p\oplus q = r\ \  \Rightarrow \ \  \left(p\oplus q\right)_4 = r_4\ \  \Rightarrow \ \  p_4 = \frac{r_4 + p\cdot q}{q_4},
\end{equation}
where we have to plug in $p_\mu$ from (\ref{eq:invert}) for the inner product $p\cdot q$. Thus we arrive at
$
p_4 = \frac{\left(q_4+r_4\right)^2}{1+q_4 r_4 - q\cdot r} -1.
$
We can summarize  the right inverse of addition as
\begin{equation}
\left[R^{-1}_q p\right]_\mu = \left(p_\mu - q_\mu\right)\frac{q_4+p_4}{1+(q_4 p_4 - q\dd p)},\qquad
\left[R^{-1}_q p\right]_4 =\frac{\left(q_4+p_4\right)^2}{1+(q_4 p_4 - q\dd p)} -1.
\end{equation}
A interesting property of this inverse is its antisymmetric property under exchange of  $p$ and $q$. 

Now we are ready to finally find the  transport operators between any two points connected by an interaction. Let us first consider 
the left transport operator $V^{p}_{k}$. It can be evaluated by inserting in (\ref{eq:leftright}) 
the expression $q=L_{p}^{-1}(k)$. We obtain

\begin{equation}
\left(V^{k}_{p}\right)^\nu_\mu = \left(p_{4}+ \frac{p\dd k}{1+ k_4}\right)\delta_\mu^\nu +\frac{p_4}{k_4\left(1+k_4\right)}k_\mu k^{\nu} - \frac{k_\mu p^{\nu}}{1+k_4}\label{eq:Vkp}
\end{equation}
We can similarly get an expression for left transport operator, this time using the right inverse 
 $\mathcal{K} =\left(q\oplus k\right)\ominus p$. Plugging in (\ref{eq:invert}) into the expression for left transport, we get



\begin{equation}
\left(U^{ k}_{p}\right)^\nu_\mu = \delta^\nu_\mu + \frac{1}{1\!+\!k_4 p_4\!-\! k\!\!\cdot\!p}\left[\frac{p_4}{k_4}\left(k_\mu k^{\nu} - p_\mu k^{\nu}\right) + k_\mu p^{\nu} - p_\mu p^{\nu}\right]\label{eq:leftU}
\end{equation}

For the inverses of all of these transport operators, see the Appendix. An interesting property of these inverses is that $(U_k^p)^{-1} = U_p^k$, but the same statement does not hold for the right transport operator.


\subsection{Parallel transport operator}
It is useful to compare the process transport operators $U$ and $V$ to the parallel transport operator given in (\ref{eq:HTrans})
in the 5 dimensional setting.
From this expression we can obtain the parallel transport operator in a local basis as 
\be
(H_{q}^{p})_{\mu}^{\nu}= e_{\mu}^{A}(q) e_{A}^{\nu}(p) -\frac{[e_{\mu}^{A}(q)P_{A}] [Q^{B}e_{B}^{\nu}(p)]}{1+P\dd Q}
\ee
Evaluating this in the embedding coordinates $P_{\mu}=p_{\mu}$ we get
\be
H_{\mu}^{\nu}= \delta_{\mu}^{\nu} + q_{\mu}p^{\nu} \left(\frac{P_{4}}{Q_{4}} - 1 \right)
-\frac{( p_{\mu}-(P\dd Q)q_{\mu})( {P_{4}}q^{\nu}-{Q_{4}}p^{\nu})}{P_{4}(1+P\dd Q)}.
\ee
This is obviously different from the expression for left transport  $U^q_r$ from Eq. (\ref{eq:leftU}).

A useful check of the expression for the left transport is to take the derivative of it and check if we get back the connection.
We might worry that taking a derivative could take us away from the constraint  $P\cdot  P =1$. However, we have 
\begin{equation}
P^A\frac{\partial}{\partial r_\nu}P_A\bigg|_{p\oplus q=r, \ r=q} =0
\end{equation}
so we should not be worried. Hence we can take take the derivative of Eq. (\ref{eq:leftU}).

%
%

A short calculation gives us that the connection defined by the left transport operator between two arbitrary momenta connected by an interaction is
\begin{equation}
\frac{\partial}{\partial r _\rho} \left(U_r^q\right)_\mu^\nu\bigg|_{q=r} = -\frac{\delta^\rho_\mu r^\nu}{r_4^2}
\end{equation}
This is not the expression for the metric compatible connection on de Sitter momentum space. It turns out however that taking the transpose of the derivative of  the left transport operator we do get 
\begin{equation}
- g^{\mu\alpha}(r)g_{\nu\beta}(r)\frac{\partial}{\partial r^\rho}\left(U_r^q\right)_\alpha^\beta\bigg|_{q=r} = g^{\mu\rho}(r) r_\nu = \Gamma^{\mu\rho}_\nu(r)
\end{equation}

so we do recover the transpose of the left transport to be metric compatible.


The above calculations lead us to the main physical result of this section: in a curved momentum space, the transport between two momenta is path dependent. Even for the left transport operator, we have that $U^p_0 U_p^q$ depends both on $q$ and $p$. The same statement holds for transports away from origin for both left and right transports. Hence there is no process-independent way of transporting between (co-)tangent spaces of two momenta $p_1$ and $p_2$. This is resulting from the fact that the addition rule is left invertible, but it is not right invertible.



\section{Closed loops of space-time events}
We will now try to find the phenomological implications of having curvature in momentum space. In the more familiar case of curved spacetime, to study the effects of curvature, one has to consider a physical process on a path in phase space, i.e. compare the results of going along path A as opposed to going along some other path B. In Relative Locality, such a process was studied before in \cite{Freidel:2011mt, Oliveira:2011gy, McCoy:2012eg} for describing the energy dependent time delay of photons coming from distant Gamma Ray Bursts. We will study a simple loop and then use the same method to comment on the Gamma Ray Burst scenario in de Sitter momentum space. 

\begin{figure}[h]
	\centering
		\includegraphics[width=0.30\textwidth]{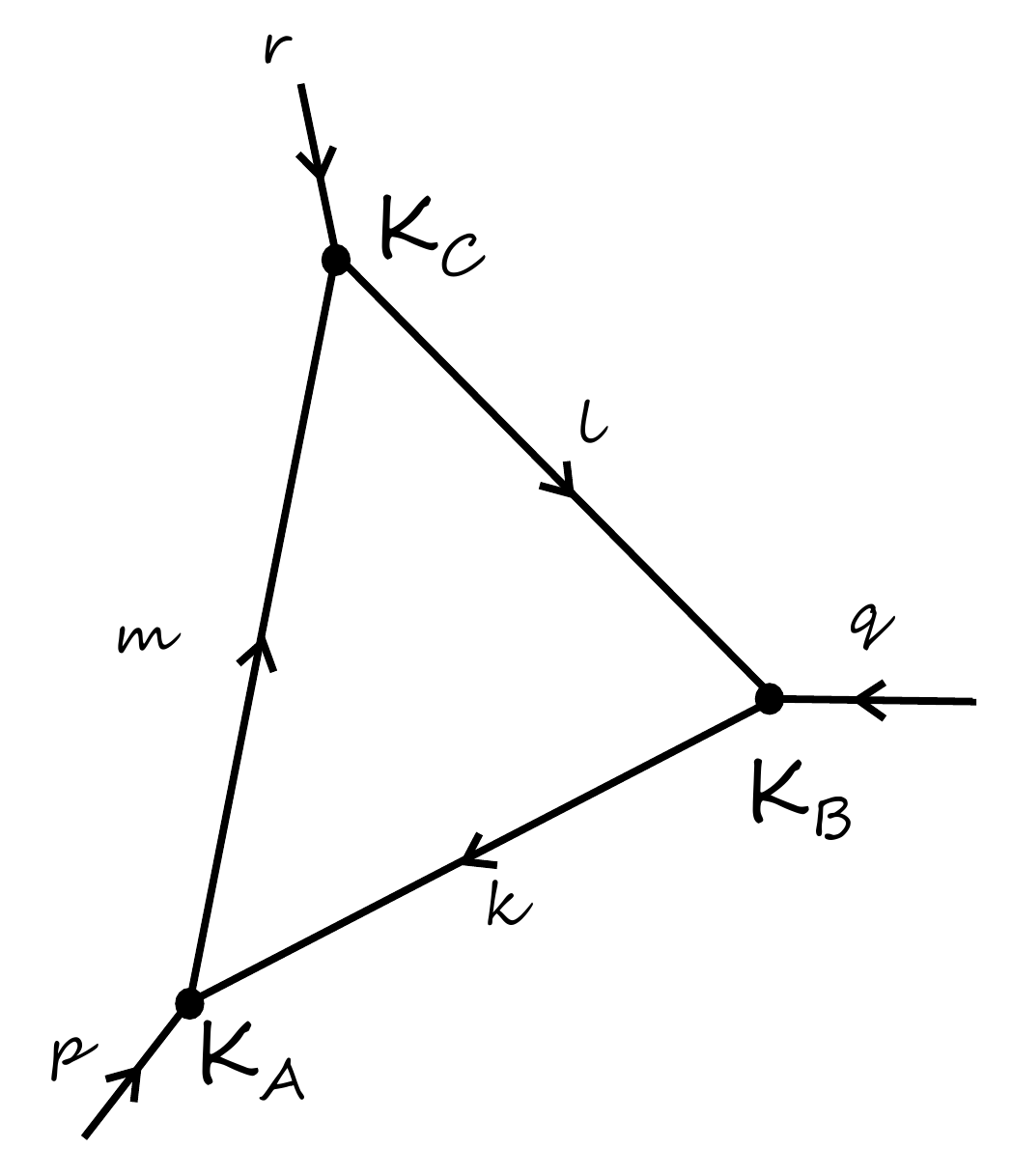}
	\caption{A simplest physical loop process}
	\label{fig:3loop}
\end{figure}

Let us start by considering a simplest loop process, with just three events. Consider a particle propagating from event A to event C for some proper time. A different particle propagates also from C, to a different point, call it B. From there another particle is sent back to A. See this in Figure \ref{fig:3loop}. For more physical processes, we can reverse some of the arrows. Since in our addition $\ominus p_\mu = -p_\mu$, this is trivial. We should be able (given the directions of the motion) to get a relation for the propagation times that closes a triangle.

Let the three conservation laws be
\begin{equation}
\mathcal{K}_A \equiv \ominus m \oplus\left(p\oplus k\right)=0, \ \ \  \mathcal{K}_B \equiv \ominus k\oplus \left(q\oplus l\right)=0, \ \ \  \mathcal{K}_C \equiv \ominus l\oplus \left(r\oplus m\right) =0.\label{eq:simplestloop}
\end{equation}
We choose these conservation laws in such a way that they define a consistent anti-clockwise orientation. Loop processes were studied  in \cite{Chen:2013fu} for the case of $\kappa$-Poincar\'e and  were shown to have global momentum conservation and other nice properties (on which we will comment later) only if the order of addition is orientable for the whole loop.

For this process to be possible in the theory, we need to not only satisfy Eq. (\ref{eq:simplestloop}), but we have to also satisfy the other equations of motion. We will show that we can write a single condition for this process to be a solution of equations of motion of the theory (plus the conservation laws). The two equations of motion that we have to keep in mind are
\begin{equation*}
x_{\textnormal{endpoint}} =\pm z\frac{\partial \mathcal{K}}{\partial p}\ \ \ \ \ \& \ \  \ \ \ \ \dot{x} = \mathcal{N}\frac{\partial D^2\left(p\right)}{\partial p}
\end{equation*}
We now have to require that the enpoints of the interactions are connected to each other. On a single worldline, this can be written for example as 
\begin{equation}
x_{l, B} = x_{l, C}+u_l \tau_l,
\end{equation}
where $x_{l,B}$ is the coordinate of the interaction $B$ on the space cotangent to momentum $l$ (similarly for interaction $C$ we have $x_{l,C}$), $u_l$ is the 4-velocity of the particle with momentum $l$ and $\tau_l$ is the propagation time. From the Appendix, we can write $u_l = \hat{l}$. For consistency we have to require that reaching point $C$ from $B$ along $l$ gives us the same point as going first to $A$ along particle $k$'s worldline and then to $C$ along worldline of momentum $m$. To write this, note that we have 
\begin{equation}
x_{l,C}=- z_C \frac{\partial \mathcal{K}_C}{\partial l} =- x_{m,C}\left(\frac{\partial \mathcal{K}_C}{\partial m}\right)^{-1}\frac{\partial \mathcal{K}_C}{\partial l}
\end{equation}
To simplify the notation, we define the following transport operators
\begin{equation}
\begin{split}
\mathcal{M}_A &= \left(\frac{\partial \mathcal{K}_A}{\partial k}\right)^{-1} \left(-\frac{\partial \mathcal{K}_A}{\partial m}\right) \\
\mathcal{M}_B &= \left(-\frac{\partial \mathcal{K}_B}{\partial k}\right)^{-1} \left(\frac{\partial \mathcal{K}_B}{\partial l}\right) \\
\mathcal{M}_C &= \left(\frac{\partial \mathcal{K}_C}{\partial m}\right)^{-1} \left(-\frac{\partial \mathcal{K}_C}{\partial l}\right), 
\end{split}
\end{equation}
the consistency now requires that we have from going in one direction in the loop that
\begin{equation}
x_{l,C} = x_{m,C}\mathcal{M}_C = \left(x_{m,A} + u_m \tau_m\right)\mathcal{M}_C = \left(x_{k,A} \mathcal{M}_A + u_m \tau_m\right)\mathcal{M}_C
\end{equation}
At the same time we have that going in the other direction yields 
\begin{equation}
x_{l,C} = x_{l,B} - u_l \tau_l = x_{k,B}\mathcal{M}_B - u_l \tau_l = \left(x_{k,A} - u_k \tau_k\right)\mathcal{M}_B - u_l \tau_l
\end{equation}
Equating the results we can thus write a single condition for the loop to be a solution of the theory, as we claimed. Simplifying, we have that the condition can be written as
\begin{equation}
\tau_l \hat{l}^\mu + \tau_k \hat{k}^\nu \mathcal{M}_{B \ \nu}^\mu + \tau_m \hat{m}^\nu \mathcal{M}_{C \ \nu}^\mu = x_{k, \ A}^\nu\left(\mathcal{M}_{B \ \nu}^\mu -   \mathcal{M}_{A \ \nu}^\rho \mathcal{M}_{C \ \rho}^\mu \right),\label{eq:3loopeqn}
\end{equation}
where $\tau$'s are propagation times of the particles, $\hat{k}, \hat{l}, \hat{m}$ are the properly normalized directions of propagation of the particles and $x_{k, \ A}^\nu$ is the coordinate of the event $A$ on the cotangent space of the particle $k$. For a  loop with different directions one just has to change the signs on the 4-velocities.

Our choice of conservation laws Eq.(\ref{eq:simplestloop}) allows us to write the $\mathcal{M}$ matrices in a simple way
\begin{equation}
\mathcal{M} = U^m_k, \ \ \ \ \mathcal{M}_B = U^l_k, \ \ \ \ \mathcal{M}_C = U^l_m
\end{equation}
Hence our right hand side of the Eq. (\ref{eq:3loopeqn}) is equal to $x_{k,A}\left( U^l_k - U^m_k U^l_m\right)$. Based on our results from the previous section, this obviously does not cancel without finetuning the momenta. One might expect that this is proportional to the curvature in momentum space. To make this statement more precise, we need to know the second derivative of the left transport operator. It can be shown that we have
\begin{equation}
\frac{\partial}{\partial p_\beta}\frac{\partial}{\partial p_\nu} \left[U_p^k\right]^\mu_\alpha - \frac{\partial}{\partial p_\mu}\frac{\partial}{\partial p_\nu} \left[U_p^k\right]^\beta_\alpha \bigg|_{p=k} = \frac{1}{2 P^2_4} R_{\ \alpha}^{\nu \ \beta\mu},
\end{equation}
and as $p\to 0$ we get the usual term for a parallel transport operator. Hence for small momenta, the loop is proportional to curvature plus higher order corrections.

 The non-cancellation of the right hand side of the equation means that depending on "where" the loop happens on the cotangent space, the relation between the propagation times is going to be different. In this sense we do not get a clean closure on the relation between the propagation times in a closed loop. This issue was pointed out in \cite{Chen:2012fu} and was called "x-dependence". In \cite{Chen:2013fu} it was shown that in $\kappa$-Poincar\'e only loops that have global momentum conservation lead to "x-independent" loops. The global momentum conservation in $\kappa$-Poincar\'e arises if and only if the order of addition is orientable for the whole loop. This statement uses crucially the properties of associativity of addition, and hence both left and right invertibility, of which the second we lack here. For example, if we check what our addition implies for total momentum conservation, we get
\begin{equation}
m = p\oplus \left(q\oplus\left(r\oplus m\right)\right).
\end{equation}
Since our addition lacks the right invertibility, this immediately gives us notrivial relation between $p, q$ and $r$, which cannot be written as simple addition. It seems that there is no way of writing the conservation laws in the non-associative case which would give us some variant of $p\oplus q\oplus r = 0$. We expect that all the loops one can write in a de Sitter momentum space with a metric connection are "x-dependent". The reason for this expectation is the non-cancellation of the simplest way of writing the conservation law, which results in the difference of products of left transport operators (which are related to the parallel transport operators). Any difference in the conservation laws is going to bring a right transport operator into the expression. Because of the nontrivial term $ \frac{p_4  + \left(\ominus k\oplus p\right)_4}{1+ k_4}\delta_\mu^\nu$ in Eq. (\ref{eq:Vkp}) for the right transport operator $V^k_p$, it is easy to see that a product of right transport operators does not cancel out (unless it is multiplied by the exact inverse).

In \cite{Freidel:2011mt, Oliveira:2011gy, McCoy:2012eg} a different loop process was considered, one in which some distant emitter sends out simultaneously two photons, one of low energy, one of high energy, which are then observed by a detector on 
Earth (or orbit -- the Fermi Gamma-ray Space Telescope). The condition for this process to be a solution in RL was written in terms of interaction coordinates $z$ as
\begin{equation}
S_2 K_2 - S_1 K_1 - T_2 P_2 + T_1 P_1 = z_1 \left(\mathcal{W}_{u_1}\mathcal{W}_{x_2}^{-1} - \mathcal{W}_{u_3}\mathcal{W}_{x_4}^{-1}\right),\label{eq:delay}
\end{equation}
In case where only nonmetricity of momentum space was present, the right hand side cancelled and it turned out that there was an energy-dependent time delay for arrival  of the two photons. For details on the setup, see \cite{Freidel:2011mt}. The right hand side in our case is proportional to $z\left(U^{p^1}_0 U^{\ominus k^1}_{p^1} \left(V^{k^1}_{p^2}\right)^{-1} U^0_{p^2} - U^{\ominus p^1}_0 V^{\ominus k^2}_{\ominus p^1} U^{\ominus p^2}_{k^2} U^0_{\ominus p^2}\right)$. Again because the transport operators are path dependent, this does not cancel. An easy way to see this is the fact that one of the terms in the equation has $V$ in it, while the other has $V^{-1}$. For this to cancel, we would require some special finetuning of momenta. Hence the calculation of time delay in a curved momentum space is observer-dependent ("x-dependent").

It thus seems that regardless of whether one interpets the different choices of $z$ as different observers, or just events happening at different places on the emergent spacetime, it is nonetheless true that there is no unique (observer-independent) way of synchronizing more than two clocks in a single process in a curved momentum space. 


\section{Discussion and conclusions}
In this work we have found a prescription for finding a momentum space manifold with action of undeformed Lorentz group preserving the metric structure and addition. The simplest solution to the resulting constraints Eq. (\ref{const}) turned out to be the de Sitter space $\mathcal{P} =  \mathrm{SO}(1,4)/\mathrm{SO}(1,3)$. It does not seem that this is the only solution, and as such it would be very interesting to clasify all the possible solutions to these constraints.

The resulting rule for addition of momenta in de Sitter has a clear geometrical interpretation -- think of two momenta $P$ and $Q$ as geodesics from the origin of $\mathcal{P}$, and construct $P\oplus Q$ as the 
geodesic starting from $P$ parallel to the $Q$ geodesic. This kind of construction can be in principle obtained for every manifold, not just a homogenous space. The only difficulty is that of solving the geodesic equation, which is however usually non-trivial.

We then went on to study the emergent spacetime structure using the Relative Locality framework. Defining interaction coordinates by parallel transport from $T^*_p\mathcal{P}\to T^*_0\mathcal{P}$, we got that the resulting spacetime has Poisson brackets reminiscent of the commutation relations of Snyder quantized spacetime. It might be interesting to try to cast in the language of Relative Locality other examples of noncommutative spaces, like the Moyal Plane \cite{2011arXiv1110.6164M} or DFR \cite{Doplicher:1994tu}.
The latter is of extra interest due to the fact that it can be obtained as a specific limit of Snyder, see \cite{Carlson:2002wj}.

In Relative Locality framework however, the transport $T^*_p\mathcal{P}\to T^*_0\mathcal{P}$ is interaction-dependent, and in general is not the parallel trasport. The process structure leading to parallel transport is also necessarily non-local. Motivated by this, we studied the different transport operators derived from the addition rule. As an application, we studied the loop processes in the theory. Since the momentum space is curved, the total ``holonomy'' around a closed loop is not identity. This results in a solution with fixed momenta and propagation times being dependent on a specific choice of $x\in T^*_p\mathcal{P}$. It has been noticed in \cite{Chen:2012fu} and requires a deeper understanding.
This phenomena is associated with configuration that forms loop in momentum space.

Note however, something similar happens in General Relativity, where parallel transport along different paths results in different answers. The process of localization is path dependent in general relativity too. The geodesics are constructed as parallel to 4-velocities, and hence momenta, so a holonomy in a closed loop is momentum dependent. In this sense, in Relative Locality we just get a dual version of this phenomenon.

\section*{Acknowledgements}
We would like to thank  Lin-Qing Chen, Florian Girelli and Lee Smolin for discussions and support. This research was supported in part by Perimeter Institute for Theoretical Physics. Research at Perimeter Institute is supported by the Government of Canada through Industry Canada and by the Province of Ontario through the Ministry of Research and Innovation.


\appendix
\section{Review of Snyder spacetime}
We review here the construction of Snyder's quantized spacetime. For more details, see the original work \cite{Snyder:1946qz}.

Snyder started from imposing
\begin{equation}
S^2 = t^2 - x^2 - y^2 - z^2\label{eq:quadform}
\end{equation}
to be an invariant of the theory, as in special relativity, but with the difference that $x$, $y$, $z$ and $t$ are Hermitian operators.
Lorentz invariance here means that the operators $x$, $y$, $z$ and $t$ have to be such that the spectra of operators $x'$, $y'$, $z'$ and $t'$,
which are linear combinations of $x$, $y$, $z$ and $t$ leaving (\ref{eq:quadform}) invariant, are the same with spectra of $x$, $y$, $z$ and $t$.

We consider the quadratic form
\begin{equation}
-\eta^2 = \eta^2_0 - \eta^2_1 - \eta^2_2 - \eta^2_3 - \eta^2_4,
\end{equation}
$\eta$'s are coordinates of a four dimensional de Sitter space $dS_4$. The operators  $x$, $y$, $z$ and $t$ can then be defined by
\begin{equation}
\begin{split}
x&=i \ell\left(\eta_4\frac{\partial}{\partial \eta_1} - \eta_1\frac{\partial}{\partial \eta_4}\right)\\
y&=i \ell\left(\eta_4\frac{\partial}{\partial \eta_2} - \eta_2\frac{\partial}{\partial \eta_4}\right)\\
z&=i \ell\left(\eta_4\frac{\partial}{\partial \eta_3} - \eta_3\frac{\partial}{\partial \eta_4}\right)\\
t&=i \ell\left(\eta_4\frac{\partial}{\partial \eta_0} + \eta_0\frac{\partial}{\partial \eta_4}\right).
\end{split}
\end{equation}
Here $\ell$ is a natural length scale, which could be the Planck length (or inverse of Planck mass). Since  $x$, $y$, $z$ 
and $t$ are Hermitian, and  $x$, $y$, $z$ "look like" angular momentum, their spectrum
is discrete with eigenvalues $m \ell$ for $m\in \mathbb{Z}_+$, whereas $t$ has a continuous spectrum. This shows that it is not necessary for 
spacetime to be continuous in order to preserve Lorentz invariance.

10 additional operators can be defined in this way. 3 boosts and 3 rotations are given by
\begin{equation}
\begin{split}
L_x&=i \hbar\left(\eta_3\frac{\partial}{\partial \eta_2} - \eta_2\frac{\partial}{\partial \eta_3}\right), \ \ \ 
M_x = i \hbar\left(\eta_0\frac{\partial}{\partial \eta_1} + \eta_1\frac{\partial}{\partial \eta_0}\right) \\
L_y&=i \hbar\left(\eta_1\frac{\partial}{\partial \eta_3} - \eta_3\frac{\partial}{\partial \eta_1}\right), \ \ \ 
M_y = i \hbar\left(\eta_0\frac{\partial}{\partial \eta_2} + \eta_2\frac{\partial}{\partial \eta_0}\right) \\
L_z&=i \hbar\left(\eta_2\frac{\partial}{\partial \eta_1} - \eta_1\frac{\partial}{\partial \eta_2}\right), \ \ \ 
M_z = i \hbar\left(\eta_0\frac{\partial}{\partial \eta_3} + \eta_3\frac{\partial}{\partial \eta_0}\right).
\end{split}
\end{equation}
Additional 4 displacement operators, energy and momentum are given by
\begin{equation}\label{eq:p}
\begin{split}
p_x&=\frac{\hbar}{\ell}\frac{\eta_1}{\eta_4}f\left(\frac{\eta_4}{\eta}\right)\ \ \ 
p_y =\frac{\hbar}{\ell}\frac{\eta_2}{\eta_4}f\left(\frac{\eta_4}{\eta}\right)\\
p_z&=\frac{\hbar}{\ell}\frac{\eta_3}{\eta_4}f\left(\frac{\eta_4}{\eta}\right)\ \ \ 
E   =\frac{\hbar}{\ell}\frac{\eta_0}{\eta_4}f\left(\frac{\eta_4}{\eta}\right),
\end{split}
\end{equation}
where $f$ is any function of $\eta_4 / \eta$. Hence we can see that for Snyder spacetime the momentum space is curved and it is the 
de Sitter space. 

A novelty of Snyder spacetime is that the position operators do not commute, for example
\begin{equation}
\left[x,y\right] = i \frac{\ell^2}{\hbar}L_z, \ \ \ \left[t,x\right]=i \frac{\ell^2}{\hbar}M_x, \ \ \ etc. \label{eq:snyder1}
\end{equation} 
which means that there would be a limit on simultaneous measurability of position coordinates. This in particular means that
the relationship between two reference frames can not be set up more precisely than allowed by these commutators. Also the commutators
of momenta and position are not the same as usual, and take the form
\begin{equation}
\left[x,p_x\right] = i\hbar\left(1+ \left(\frac{\ell}{\hbar}\right)^2p_x^2\right), \ \ \ 
\left[x,p_y\right] = i\hbar\left(\frac{\ell}{\hbar}\right)^2p_xp_y, \ \ \ etc. \label{eq:snyder2}
\end{equation}
We can see that the usual commutator between momentum and position has an additional term depending on the length scale. 
In the limit of $\ell\rightarrow 0$ with $\hbar$ fixed, all of these commutation relations become the familiar ones. 

The work towards Snyder's original goal of formulating a quantum field theory on this space is still ongoing, see for example
\cite{Battisti:2010sr, Girelli:2010wi}.  Snyder's work turned out to be very influential, as it spawned a whole field of 
quantum noncommutative spacetimes, which was revived in 1995 in form
of DFR spacetimes \cite{Doplicher:1994tu}. It has also been studied in the framework of Deformed Special Relativity, see for
example \cite{Girelli:2004ue}. 


\section{Inverses of transport operators}
In this appendix we show how to invert the transport operators $U$ and $V$. We know that $U$ and $V$ are elements of the Lorentz group, and as such, we can write them as
\begin{equation}
U = a\delta_\mu^\nu + b k_\mu k^\nu + c p_\mu p^\nu + d k_\mu p^\nu + e p_\mu k^\nu,
\end{equation}
where $a,\ldots e$ are functions of the momenta $k$ and $p$; similarly we can write $V$. As such, the inverse can then be written as
\begin{equation}
U^{-1} = u\delta_\mu^\nu + v k_\mu k^\nu + w p_\mu p^\nu + x k_\mu p^\nu + y p_\mu k^\nu,
\end{equation}
with $u,\ldots y$ again functions of the two momenta. To solve for the unknown functions $u,\ldots y$ we require that $U U^{-1} = \mathds{1}$. This gives us
\begin{equation}
\left\{ \begin{array}{l}
u = \frac{1}{a}\\ u b + v\left(a + b k\!\cdot\! k + d k\!\cdot\! p\right) + y\left(b k\!\cdot\! p + d p \!\cdot\! p\right) = 0  \\
u c + w\left(a + c p\!\cdot\! p + e k \!\cdot\! p\right) + x\left(c k\!\cdot\! p + e k\!\cdot\! k\right) = 0 \\
u d + w \left(b k\!\cdot\! p + d p\!\cdot\! p\right) + x\left(a + b k\!\cdot\! k + d k\!\cdot\! p\right) = 0 \\
u e + v\left(c k\!\cdot\! p + e k\!\cdot\! k\right) + y\left(a + c p\!\cdot\! p + e k\!\cdot\! p\right) = 0 \label{eq:nonsimpleeqns}
\end{array}\right.
\end{equation}
We notice that we can simplify the equations a bit if we denote
\begin{equation}
\begin{split}
f \equiv a + b k\!\cdot\! k + d k\!\cdot\! p, \ \ \ g \equiv b k\!\cdot\! p + d p \!\cdot\! p, \\
h \equiv a + c p\!\cdot\! p + e k \!\cdot\! p, \ \ \  i \equiv c k\!\cdot\! p + e k\!\cdot\! k.
\end{split}
\end{equation}

We can now readily solve this system of equations to get
\begin{equation}
\left\{ \begin{array}{l}
u = \frac{1}{a}\\
v = -u\left(\frac{b}{f} + \frac{g}{f}\frac{ef -bi}{gi - fh}\right) \\
w = -u\left(\frac{d}{g} + \frac{f}{g}\frac{hd - cg}{gi - fh}\right) \\
x = u\frac{hd - cg}{gi - fh}\\
y = u \frac{ef -bi}{gi - fh}
\end{array}\right.
\end{equation}

Note however, that in some cases it is necessary to use the unsimplified Eq. (\ref{eq:nonsimpleeqns}), as some of the terms might be 0.

Let us now consider the inverses of the operators $U$ and $V$.  We find that it is actually trivial to solve for the inverses between the origin and some point, in which case we get
\begin{align}
\left[\left(U^0_{p}\right)^{-1}\right]^\nu_\mu &=  \left(U^p_{0}\right)^\nu_\mu =  \delta^\nu_\mu + \frac{p_\mu p^\nu}{P_4(1+P_4)} \\
\left[\left(V^0_{q}\right)^{-1}\right]^\nu_\mu &= \frac{\delta^\nu_\mu}{Q_4} \\
\left[\left(V^p_{0}\right)^{-1}\right]^\nu_\mu &= \left(U^0_{p}\right)^\nu_\mu 
\end{align}

For the inverse of the left transport between $q$ and $p\oplus q$ we get
\begin{equation}
\begin{split}
\left[\left(U^q_{p\oplus q}\right)^{-1}\right]^\nu_\mu &= \delta^\nu_\mu + \frac{1}{\left(1+P_4\right)\left(P_4 - \frac{p\cdot q}{Q_4}\right)}p_\mu p^\nu + \frac{1}{Q_4\left(P_4 - \frac{p\cdot q}{Q_4}\right)}p_\mu q^\nu \\
&= \delta^\nu_\mu + \frac{Q_4}{\left(1+P_4\right)\left(P\oplus Q\right)_4}p_\mu p^\nu + \frac{1}{\left(P\oplus Q\right)_4}p_\mu q^\nu \label{eq:Uinverse}
\end{split}
\end{equation}
We can plug in now a conservation law $q\oplus k = p$ to get the inverse left transport 
\begin{equation}
\left[\left(U^{k}_{p}\right)^{-1}\right]^\nu_\mu = \delta^\nu_\mu + \frac{1}{1\!+\!k_4 p_4\!-\! k\!\!\cdot\!p}\left[\frac{k_4}{p_4}\left(p_\mu p^{\nu} - k_\mu p^{\nu}\right) + p_\mu k^{\nu} - k_\mu k^{\nu}\right]
\end{equation}
Hence our claim from the main text that $(U_k^p)^{-1} = U_p^k$ is proved.

Similarly we can solve for the inverse of the right transport operator to get

\begin{equation}
\left[\left(V^p_{p\oplus q}\right)^{-1}\right]^\nu_\mu =\frac{1+P_4}{Q_4 + \left(P\oplus Q\right)_4}\delta^\nu_\mu + \frac{1}{\left(Q_4 + \left(P\oplus Q\right)_4\right)\left(P\oplus Q\right)_4}\left(\frac{p\!\cdot\! q}{\left(1+P_4\right)}p_\mu p^\nu + P_4 p_\mu q^\nu\right) \label{eq:Vinverse}
\end{equation}
Note that it is much easier this way to find the inverse than inverting explicitly the expression (\ref{eq:Vkp}), since we have for $p\oplus q = r$ that $Q_4 = \left(\ominus P\oplus R\right)_4$. Using (\ref{eq:invert}) on (\ref{eq:Uinverse}) and (\ref{eq:Vinverse}) we can now get all the expressions we need.

With this in mind, we write down $\left(V^{k}_{k\ominus r}\right)^{-1}$ and impose $p = k\ominus r$. A short calculation gives us
\begin{equation}
\begin{split}
\left[\left(V^{k}_{p}\right)^{-1}\right]^\nu_\mu &= \frac{1+k_4}{p_4 + r_4}\delta^\nu_\mu - \frac{1}{p_4\left(1+k_4\right)}\left(1 - \frac{k\!\cdot\! p}{p_4 +r_4}\right)k_\mu k^{ \nu} + \frac{k_4}{p_4\left(p_4 + r_4\right)}k_\mu p^{\nu} \\
r_4 &= \left(\ominus k\oplus p\right)_4
\end{split}
\end{equation}
It is an immediate result that in general $(V_k^p)^{-1} \neq V_p^k$.


\section{Massless particles}
An interesting simplification occurs in the Snyder momentum space when one considers a theory of only massless particles. For each particle then the 4th component is $P_4=1$. With this the addition simply becomes
\begin{equation}
\begin{split}
(p\oplus q)_4 &= 1 - p\cdot q \\
(p\oplus q)_\mu &= q_\mu + p_\mu\frac{2 - p\cdot q}{2}
\end{split}
\end{equation}
We can immediately see that restricting our attention to 3-vertex interactions, we necessarily get $p\cdot q = 0$, which means that the addition becomes that of a flat Minkowski. In this sense only massive particles are exposed to the effects of curvature in Snyder momentum space. 

Note however, that the transport operators are still not trivial in this limit, and they are given by
\begin{equation}
\begin{split}
\left(U^0_{p}\right)^\nu_\mu &= \delta^\nu_\mu - \frac{p_\mu p^\nu}{2} \\
\left(V^0_{q}\right)^\nu_\mu &=  \delta^\nu_\mu \\
\left(V^p_{0}\right)^\nu_\mu &= \left(U^p_{0}\right)^\nu_\mu =  \delta^\nu_\mu + \frac{p_\mu p^\nu}{2} \\
\left(U^{ k}_{p}\right)^\nu_\mu &= \delta^\nu_\mu + \frac{k_\mu k^{\nu} - p_\mu k^{\nu} + k_\mu p^{\nu} - p_\mu p^{\nu}}{2} \\
\left(V^{k}_{p}\right)^\nu_\mu &= \delta_\mu^\nu -\frac{k_\mu k^{\nu}}{2} + \frac{k_\mu p^{\nu}}{2}
\end{split}
\end{equation}
It is interesting to note that even in this limit $U_0^p U_p^k$ is still dependent on $p$.


\section{4-velocity}
In the 4d coordinates we use our expression for the geodesics to get
\begin{equation}
D(p) = \cosh^{-1}\left(p_4\right).
\end{equation}
This is an obvious result, since we have already found the expression in the embedding coordinates, and the geodesic distance is a diffeomorphism invariant quantity.
Using the equation of motion, this gives us the velocity to be
\begin{equation}
\dot{x}^\mu = \mathcal{N}\frac{\partial D^2(p)}{\partial p_\mu} = -\mathcal{N}\frac{\cosh^{-1}\left(p_4\right)}{p_4\sqrt{p_4^2-1}}p^\mu.
\end{equation}
Hence we get a momentum dependant factor in front of every velocity one would calculate in the Riemann Normal Coordinates, which does not make for much trouble. Additionaly, we know that actually $p_4$ is related to mass by $p_4 = \cosh m$, so it is just an expression involving mass. Requiring that $\dot{x}^2 = -1$ and that $\dot{x}^\mu$ has the same direction as the momentum, we get that
\begin{equation}
 \mathcal{N} = - \frac{p_4}{\cosh^{-1}\left(p_4\right)},
\end{equation}
so that for massive particles we have
\begin{equation}
\dot{x}^\mu = \frac{p^\mu}{\sqrt{p_4^2-1}}.
\end{equation}

We also notice that for massless particles we would seemingly divide by 0, but being careful in taking the limit we notice that
\begin{equation}
\lim_{p_4\rightarrow1} \frac{\cosh^{-1}\left(p_4\right)}{p_4\sqrt{p_4^2-1}} = 1.
\end{equation}


\end{document}